\catcode`@=11 

\font\fourteenrm=cmr10 scaled\magstep2
\font\twelverm=cmr10 scaled\magstep1
\font\ninerm=cmr9            \font\sixrm=cmr6

\font\fourteenbf=cmbx10 scaled\magstep2
\font\twelvebf=cmbx10 scaled\magstep1
\font\ninebf=cmbx9            \font\sixbf=cmbx6
\font\seventeeni=cmmi10 scaled\magstep3     \skewchar\seventeeni='177
\font\fourteeni=cmmi10 scaled\magstep2      \skewchar\fourteeni='177
\font\twelvei=cmmi10 scaled\magstep1        \skewchar\twelvei='177
\font\ninei=cmmi9                           \skewchar\ninei='177
\font\sixi=cmmi6                            \skewchar\sixi='177
\font\seventeensy=cmsy10 scaled\magstep3    \skewchar\seventeensy='60
\font\fourteensy=cmsy10 scaled\magstep2     \skewchar\fourteensy='60
\font\twelvesy=cmsy10 scaled\magstep1       \skewchar\twelvesy='60
\font\ninesy=cmsy9                          \skewchar\ninesy='60
\font\sixsy=cmsy6                           \skewchar\sixsy='60

\font\fourteenex=cmex10 scaled\magstep2
\font\twelveex=cmex10 scaled\magstep1

\font\fourteensl=cmsl10 scaled\magstep2
\font\twelvesl=cmsl10 scaled\magstep1
\font\ninesl=cmsl9

\font\fourteenit=cmti10 scaled\magstep2
\font\twelveit=cmti10 scaled\magstep1
\font\twelvett=cmtt10 scaled\magstep1
\font\twelvecp=cmcsc10 scaled\magstep1
\font\tencp=cmcsc10
\newfam\cpfam
%
%
\newcount\f@ntkey            \f@ntkey=0
\def\samef@nt{\relax \ifcase\f@ntkey \rm \or\oldstyle \or\or
         \or\it \or\sl \or\bf \or\tt \or\caps \fi }
\def\fourteenpoint{\relax
    \textfont0=\fourteenrm          \scriptfont0=\tenrm
    \scriptscriptfont0=\sevenrm
     \def\rm{\fam0 \fourteenrm \f@ntkey=0 }\relax
    \textfont1=\fourteeni           \scriptfont1=\teni
    \scriptscriptfont1=\seveni
     \def\oldstyle{\fam1 \fourteeni\f@ntkey=1 }\relax
    \textfont2=\fourteensy          \scriptfont2=\tensy
    \scriptscriptfont2=\sevensy
    \textfont3=\fourteenex     \scriptfont3=\fourteenex
    \scriptscriptfont3=\fourteenex
    \def\it{\fam\itfam \fourteenit\f@ntkey=4 }\textfont\itfam=\fourteenit
    \def\sl{\fam\slfam \fourteensl\f@ntkey=5 }\textfont\slfam=\fourteensl
    \scriptfont\slfam=\tensl
    \def\bf{\fam\bffam \fourteenbf\f@ntkey=6 }\textfont\bffam=\fourteenbf
    \scriptfont\bffam=\tenbf     \scriptscriptfont\bffam=\sevenbf
    \def\tt{\fam\ttfam \twelvett \f@ntkey=7 }\textfont\ttfam=\twelvett
    \h@big=11.9\p@{} \h@Big=16.1\p@{} \h@bigg=20.3\p@{} \h@Bigg=24.5\p@{}
    \def\caps{\fam\cpfam \twelvecp \f@ntkey=8 }\textfont\cpfam=\twelvecp
    \setbox\strutbox=\hbox{\vrule height 12pt depth 5pt width\z@}
    \samef@nt}
\def\twelvepoint{\relax
    \textfont0=\twelverm          \scriptfont0=\ninerm
    \scriptscriptfont0=\sixrm
     \def\rm{\fam0 \twelverm \f@ntkey=0 }\relax
    \textfont1=\twelvei           \scriptfont1=\ninei
    \scriptscriptfont1=\sixi
     \def\oldstyle{\fam1 \twelvei\f@ntkey=1 }\relax
    \textfont2=\twelvesy          \scriptfont2=\ninesy
    \scriptscriptfont2=\sixsy
    \textfont3=\twelveex          \scriptfont3=\twelveex
    \scriptscriptfont3=\twelveex
    \def\it{\fam\itfam \twelveit \f@ntkey=4 }\textfont\itfam=\twelveit
    \def\sl{\fam\slfam \twelvesl \f@ntkey=5 }\textfont\slfam=\twelvesl
    \scriptfont\slfam=\ninesl
    \def\bf{\fam\bffam \twelvebf \f@ntkey=6 }\textfont\bffam=\twelvebf
    \scriptfont\bffam=\ninebf     \scriptscriptfont\bffam=\sixbf
    \def\tt{\fam\ttfam \twelvett \f@ntkey=7 }\textfont\ttfam=\twelvett
    \h@big=10.2\p@{}
    \h@Big=13.8\p@{}
    \h@bigg=17.4\p@{}
    \h@Bigg=21.0\p@{}
    \def\caps{\fam\cpfam \twelvecp \f@ntkey=8 }\textfont\cpfam=\twelvecp
    \setbox\strutbox=\hbox{\vrule height 10pt depth 4pt width\z@}
    \samef@nt}
\def\tenpoint{\relax
    \textfont0=\tenrm          \scriptfont0=\sevenrm
    \scriptscriptfont0=\fiverm
    \def\rm{\fam0 \tenrm \f@ntkey=0 }\relax
    \textfont1=\teni           \scriptfont1=\seveni
    \scriptscriptfont1=\fivei
    \def\oldstyle{\fam1 \teni \f@ntkey=1 }\relax
    \textfont2=\tensy          \scriptfont2=\sevensy
    \scriptscriptfont2=\fivesy
    \textfont3=\tenex          \scriptfont3=\tenex
    \scriptscriptfont3=\tenex
    \def\it{\fam\itfam \tenit \f@ntkey=4 }\textfont\itfam=\tenit
    \def\sl{\fam\slfam \tensl \f@ntkey=5 }\textfont\slfam=\tensl
    \def\bf{\fam\bffam \tenbf \f@ntkey=6 }\textfont\bffam=\tenbf
    \scriptfont\bffam=\sevenbf     \scriptscriptfont\bffam=\fivebf
    \def\tt{\fam\ttfam \tentt \f@ntkey=7 }\textfont\ttfam=\tentt
    \def\caps{\fam\cpfam \tencp \f@ntkey=8 }\textfont\cpfam=\tencp
    \setbox\strutbox=\hbox{\vrule height 8.5pt depth 3.5pt width\z@}
    \samef@nt}
%
%
\newdimen\h@big  \h@big=8.5\p@
\newdimen\h@Big  \h@Big=11.5\p@
\newdimen\h@bigg  \h@bigg=14.5\p@
\newdimen\h@Bigg  \h@Bigg=17.5\p@
\def\big#1{{\hbox{$\left#1\vbox to\h@big{}\right.\n@space$}}}
\def\Big#1{{\hbox{$\left#1\vbox to\h@Big{}\right.\n@space$}}}
\def\bigg#1{{\hbox{$\left#1\vbox to\h@bigg{}\right.\n@space$}}}
\def\Bigg#1{{\hbox{$\left#1\vbox to\h@Bigg{}\right.\n@space$}}}
%
%
\normalbaselineskip = 20pt plus 0.2pt minus 0.1pt
\normallineskip = 0pt 
\normallineskiplimit = 0pt
\newskip\normaldisplayskip
\normaldisplayskip = 8pt plus 5pt minus 3pt
\newskip\normaldispshortskip
\normaldispshortskip = 6pt plus 2pt
\newskip\normalparskip
\normalparskip = 3pt
\newskip\skipregister
\skipregister = 4pt plus 2pt minus .5pt
\newif\ifsingl@    \newif\ifdoubl@
\newif\iftwelv@    \twelv@true
\def\singlespace{\singl@true\doubl@false\spaces@t}
\def\doublespace{\singl@false\doubl@true\spaces@t}
\def\normalspace{\singl@false\doubl@false\spaces@t}
\def\Tenpoint{\tenpoint\twelv@false\spaces@t}
\def\Twelvepoint{\twelvepoint\twelv@true\spaces@t}
\def\spaces@t{\relax%
 \iftwelv@ \ifsingl@\subspaces@t3:4;\else\subspaces@t1:1;\fi%
 \else \ifsingl@\subspaces@t3:5;\else\subspaces@t4:5;\fi \fi%
 \ifdoubl@ \multiply\baselineskip by 5%
 \divide\baselineskip by 4 \fi \unskip}
\def\subspaces@t#1:#2;{
      \baselineskip = \normalbaselineskip
      \multiply\baselineskip by #1 \divide\baselineskip by #2
      \lineskip = \normallineskip
      \multiply\lineskip by #1 \divide\lineskip by #2
      \lineskiplimit = \normallineskiplimit
      \multiply\lineskiplimit by #1 \divide\lineskiplimit by #2
      \parskip = \normalparskip
      \multiply\parskip by #1 \divide\parskip by #2
      \abovedisplayskip = \normaldisplayskip
      \multiply\abovedisplayskip by #1 \divide\abovedisplayskip by #2
      \belowdisplayskip = \abovedisplayskip
      \abovedisplayshortskip = \normaldispshortskip
      \multiply\abovedisplayshortskip by #1
        \divide\abovedisplayshortskip by #2
      \belowdisplayshortskip = \abovedisplayshortskip
      \advance\belowdisplayshortskip by \belowdisplayskip
      \divide\belowdisplayshortskip by 2
      \smallskipamount = \skipregister
      \multiply\smallskipamount by #1 \divide\smallskipamount by #2
      \medskipamount = \smallskipamount \multiply\medskipamount by 2
      \bigskipamount = \smallskipamount \multiply\bigskipamount by 4 }
\def\normalbaselines{ \baselineskip=\normalbaselineskip
   \lineskip=\normallineskip \lineskiplimit=\normallineskip
   \iftwelv@\else \multiply\baselineskip by 4 \divide\baselineskip by 5
     \multiply\lineskiplimit by 4 \divide\lineskiplimit by 5
     \multiply\lineskip by 4 \divide\lineskip by 5 \fi }
\Twelvepoint  
\interlinepenalty=50
\interfootnotelinepenalty=5000
\predisplaypenalty=9000
\postdisplaypenalty=500
\hfuzz=1pt
\vfuzz=0.2pt
%
%
\def\pagecontents{
   \ifvoid\topins\else\unvbox\topins\vskip\skip\topins\fi
   \dimen@ = \dp255 \unvbox255
   \ifvoid\footins\else\vskip\skip\footins\footrule\unvbox\footins\fi
   \ifr@ggedbottom \kern-\dimen@ \vfil \fi }
\def\makeheadline{\vbox to 0pt{ \skip@=\topskip
      \advance\skip@ by -12pt \advance\skip@ by -2\normalbaselineskip
      \vskip\skip@ \line{\vbox to 12pt{}\the\headline} \vss
      }\nointerlineskip}
\def\makefootline{\baselineskip = 1.5\normalbaselineskip
                 \line{\the\footline}}
\newif\iffrontpage
\newif\ifletterstyle
\newif\ifp@genum
\def\nopagenumbers{\p@genumfalse}
\def\pagenumbers{\p@genumtrue}
\pagenumbers
\newtoks\paperheadline
\newtoks\letterheadline
\newtoks\letterfrontheadline
\newtoks\lettermainheadline
\newtoks\paperfootline
\newtoks\letterfootline
\newtoks\date
\footline={\the\paperfootline}
\paperfootline={\hss\iffrontpage\else\ifp@genum\tenrm\folio\hss\fi\fi}
\letterfootline={\hfil}
\headline={\the\paperheadline}
\paperheadline={\hfil}
\countdef\pagenumber=1  \pagenumber=1
\def\advancepageno{\global\advance\pageno by 1
   \ifnum\pagenumber<0 \global\advance\pagenumber by -1
    \else\global\advance\pagenumber by 1 \fi \global\frontpagefalse }
\def\folio{\ifnum\pagenumber<0 \romannumeral-\pagenumber
           \else \number\pagenumber \fi }
\def\footrule{\dimen@=\prevdepth\nointerlineskip
   \vbox to 0pt{\vskip -0.25\baselineskip \hrule width 0.35\hsize \vss}
   \prevdepth=\dimen@ }
\newtoks\foottokens
\foottokens={\Tenpoint\singlespace}
\newdimen\footindent
\footindent=24pt
\def\vfootnote#1{\insert\footins\bgroup  \the\foottokens
   \interlinepenalty=\interfootnotelinepenalty \floatingpenalty=20000
   \splittopskip=\ht\strutbox \boxmaxdepth=\dp\strutbox
   \leftskip=\footindent \rightskip=\z@skip
   \parindent=0.5\footindent \parfillskip=0pt plus 1fil
   \spaceskip=\z@skip \xspaceskip=\z@skip
   \Textindent{$ #1 $}\footstrut\futurelet\next\fo@t}
\def\Textindent#1{\noindent\llap{#1\enspace}\ignorespaces}
\def\footnote#1{\attach{#1}\vfootnote{#1}}

\let\footsymbol=\star
\newcount\lastf@@t           \lastf@@t=-1
\newcount\footsymbolcount    \footsymbolcount=0
\newif\ifPhysRev
\def\footsymbolgen{\relax \ifPhysRev \iffrontpage \NPsymbolgen\else
      \PRsymbolgen\fi \else \NPsymbolgen\fi
   \global\lastf@@t=\pageno \footsymbol }
\def\NPsymbolgen{\ifnum\footsymbolcount<0 \global\footsymbolcount=0\fi
   {\iffrontpage \else \advance\lastf@@t by 1 \fi
    \ifnum\lastf@@t<\pageno \global\footsymbolcount=0
     \else \global\advance\footsymbolcount by 1 \fi }
   \ifcase\footsymbolcount \fd@f\star\or \fd@f\dagger\or \fd@f\ast\or
    \fd@f\ddagger\or \fd@f\natural\or \fd@f\diamond\or \fd@f\bullet\or
    \fd@f\nabla\else \fd@f\dagger\global\footsymbolcount=0 \fi }
\def\fd@f#1{\xdef\footsymbol{#1}}
\def\PRsymbolgen{\ifnum\footsymbolcount>0 \global\footsymbolcount=0\fi
      \global\advance\footsymbolcount by -1
      \xdef\footsymbol{\sharp\number-\footsymbolcount} }
\def\space@ver#1{\let\@sf=\empty \ifmmode #1\else \ifhmode
   \edef\@sf{\spacefactor=\the\spacefactor}\unskip${}#1$\relax\fi\fi}
\newcount\chapternumber      \chapternumber=0
\newcount\sectionnumber      \sectionnumber=0
\newcount\equanumber         \equanumber=0
\let\chapterlabel=0
\newtoks\chapterstyle        \chapterstyle={\Number}
\newskip\chapterskip         \chapterskip=\bigskipamount
\newskip\sectionskip         \sectionskip=\medskipamount
\newskip\headskip            \headskip=6pt plus 1pt
\newdimen\chapterminspace    \chapterminspace=5pc
\newdimen\sectionminspace    \sectionminspace=3pc
\newdimen\referenceminspace  \referenceminspace=5pc
\def\chapterreset{\global\advance\chapternumber by 1
   \ifnum\the\equanumber<0 \else\global\equanumber=0\fi
   \sectionnumber=0 \makel@bel}
\def\makel@bel{\xdef\chapterlabel{%
\the\chapterstyle{\the\chapternumber}.}}
\def\sectionlabel{\number\sectionnumber \quad }
\def\alphabetic#1{\count255='140 \advance\count255 by #1\char\count255}
\def\Alphabetic#1{\count255='100 \advance\count255 by #1\char\count255}
\def\Roman#1{\uppercase\expandafter{\romannumeral #1}}
\def\roman#1{\romannumeral #1}
\def\Number#1{\number #1}
\def\unnumberedchapters{\let\makel@bel=\relax \let\chapterlabel=\relax
\let\sectionlabel=\relax \equanumber=-1 }
\def\titlestyle#1{\par\begingroup \interlinepenalty=9999
     \leftskip=0.02\hsize plus 0.23\hsize minus 0.02\hsize
     \rightskip=\leftskip \parfillskip=0pt
     \hyphenpenalty=9000 \exhyphenpenalty=9000
     \tolerance=9999 \pretolerance=9000
     \spaceskip=0.333em \xspaceskip=0.5em
     \iftwelv@\fourteenpoint\else\twelvepoint\fi
   \noindent #1\par\endgroup }
\def\spacecheck#1{\dimen@=\pagegoal\advance\dimen@ by -\pagetotal
   \ifdim\dimen@<#1 \ifdim\dimen@>0pt \vfil\break \fi\fi}
\def\chapter#1{\par \penalty-300 \vskip\chapterskip
   \spacecheck\chapterminspace
   \chapterreset \titlestyle{\chapterlabel \ #1}
   \nobreak\vskip\headskip \penalty 30000
   \wlog{\string\chapter\ \chapterlabel} }

\def\section#1{\par \ifnum\the\lastpenalty=30000\else
   \penalty-200\vskip\sectionskip \spacecheck\sectionminspace\fi
   \wlog{\string\section\ \chapterlabel \the\sectionnumber}
   \global\advance\sectionnumber by 1  \noindent
   {\caps\enspace\chapterlabel \sectionlabel #1}\par
   \nobreak\vskip\headskip \penalty 30000 }
\def\subsection#1{\par
   \ifnum\the\lastpenalty=30000\else \penalty-100\smallskip \fi
   \noindent\undertext{#1}\enspace \vadjust{\penalty5000}}

\def\undertext#1{\vtop{\hbox{#1}\kern 1pt \hrule}}
\def\APPENDIX#1#2{\par\penalty-300\vskip\chapterskip
   \spacecheck\chapterminspace \chapterreset \xdef\chapterlabel{#1}
   \titlestyle{APPENDIX #2} \nobreak\vskip\headskip \penalty 30000
   \wlog{\string\Appendix\ \chapterlabel} }
\def\Appendix#1{\APPENDIX{#1}{#1}}
\def\appendix{\APPENDIX{A}{}}
\def\eqname#1{\relax \ifnum\the\equanumber<0%
     \xdef#1{{\noexpand\rm(\number-\equanumber)}}%
     \global\advance\equanumber by -1%
    \else \global\advance\equanumber by 1%
      \xdef#1{{\noexpand\rm(\chapterlabel \number\equanumber)}} \fi}
\def\eqinsert#1{\noalign{\dimen@=\prevdepth \nointerlineskip
   \setbox0=\hbox to\displaywidth{\hfil #1}
   \vbox to 0pt{\vss\hbox{$\!\box0\!$}\kern-0.5\baselineskip}
   \prevdepth=\dimen@}}
\def\GENITEM#1;#2{\par \hangafter=0 \hangindent=#1
    \Textindent{$ #2 $}\ignorespaces}
\outer\def\newitem#1=#2;{\gdef#1{\GENITEM #2;}}
\newdimen\itemsize                \itemsize=30pt
\newitem\item=1\itemsize;
\newitem\sitem=1.75\itemsize;     
\newitem\ssitem=2.5\itemsize;     
\outer\def\newlist#1=#2&#3&#4;{\toks0={#2}\toks1={#3}%
   \count255=\escapechar \escapechar=-1
   \alloc@0\list\countdef\insc@unt\listcount     \listcount=0
   \edef#1{\par
      \countdef\listcount=\the\allocationnumber
      \advance\listcount by 1
      \hangafter=0 \hangindent=#4
      \Textindent{\the\toks0{\listcount}\the\toks1}}
   \expandafter\expandafter\expandafter
    \edef\c@t#1{begin}{\par
      \countdef\listcount=\the\allocationnumber \listcount=1
      \hangafter=0 \hangindent=#4
      \Textindent{\the\toks0{\listcount}\the\toks1}}
   \expandafter\expandafter\expandafter
    \edef\c@t#1{con}{\par \hangafter=0 \hangindent=#4 \noindent}
   \escapechar=\count255}
\def\c@t#1#2{\csname\string#1#2\endcsname}
\newlist\point=\Number&.&1.0\itemsize;
\newlist\subpoint=(\alphabetic&)&1.75\itemsize;
\newlist\subsubpoint=(\roman&)&2.5\itemsize;
\newcount\referencecount     \referencecount=0
\newif\ifreferenceopen       \newwrite\referencewrite
\newtoks\rw@toks
\newcount\lastrefsbegincount \lastrefsbegincount=0
\def\refch@ck{\chardef\rw@write=\referencewrite
   \ifreferenceopen \else \referenceopentrue
   \immediate\openout\referencewrite=referenc.texauxil \fi}
%
{\catcode`\^^M=\active 
  \gdef\obeyendofline{\catcode`\^^M\active \let^^M\ }}%
%
{\catcode`\^^M=\active 
  \gdef\ignoreendofline{\catcode`\^^M=5}}
{\obeyendofline\gdef\rw@start#1{\def\t@st{#1} \ifx\t@st\blankend%
\endgroup \@sf \relax \else \ifx\t@st\bl@nkend \endgroup \@sf \relax%
\else \rw@begin#1
\backtotext
\fi \fi } }
{\obeyendofline\gdef\rw@begin#1
{\def\n@xt{#1}\rw@toks={#1}\relax%
\rw@next}}
\def\blankend{}
{\obeylines\gdef\bl@nkend{
}}
\newif\iffirstrefline  \firstreflinetrue
\def\rwr@teswitch{\ifx\n@xt\blankend \let\n@xt=\rw@begin %
 \else\iffirstrefline \global\firstreflinefalse%
\immediate\write\rw@write{\noexpand\obeyendofline \the\rw@toks}%
\let\n@xt=\rw@begin%
      \else\ifx\n@xt\rw@@d \def\n@xt{\immediate\write\rw@write{%
        \noexpand\ignoreendofline}\endgroup \@sf}%
             \else \immediate\write\rw@write{\the\rw@toks}%
             \let\n@xt=\rw@begin\fi\fi \fi}
\def\rw@next{\rwr@teswitch\n@xt}
\def\rw@@d{\backtotext} \let\rw@end=\relax
\let\backtotext=\relax

\newdimen\refindent     \refindent=30pt
\def\refitem#1{\par \hangafter=0 \hangindent=\refindent \Textindent{#1}}
\def\REFNUM#1{\space@ver{}\refch@ck \firstreflinetrue%
 \global\advance\referencecount by 1 \xdef#1{\the\referencecount}}
\def\refnum#1{\space@ver{}\refch@ck \firstreflinetrue%
 \global\advance\referencecount by 1 \xdef#1{\the\referencecount}\refend}

\def\REF#1{\REFNUM#1%
 \immediate\write\referencewrite{%
 \noexpand\refitem{#1.}}%
\begingroup\obeyendofline\rw@start}
\def\ref{\refnum\?%
 \immediate\write\referencewrite{\noexpand\refitem{\?.}}%
\begingroup\obeyendofline\rw@start}
\def\Ref#1{\refnum#1%
 \immediate\write\referencewrite{\noexpand\refitem{#1.}}%
\begingroup\obeyendofline\rw@start}
\def\REFS#1{\REFNUM#1\global\lastrefsbegincount=\referencecount
\immediate\write\referencewrite{\noexpand\refitem{#1.}}%
\begingroup\obeyendofline\rw@start}

\def\REFSCON#1{\REF#1}
\newcount\figurecount     \figurecount=0
\newif\iffigureopen       \newwrite\figurewrite
\def\figch@ck{\chardef\rw@write=\figurewrite \iffigureopen\else
   \immediate\openout\figurewrite=figures.texauxil
   \figureopentrue\fi}
\def\FIGNUM#1{\space@ver{}\figch@ck \firstreflinetrue%
 \global\advance\figurecount by 1 \xdef#1{\the\figurecount}}
\def\FIG#1{\FIGNUM#1
   \immediate\write\figurewrite{\noexpand\refitem{#1.}}%
   \begingroup\obeyendofline\rw@start}
\def\fig{\FIGNUM\? fig. \?%
\immediate\write\figurewrite{\noexpand\refitem{\?.}}%
\begingroup\obeyendofline\rw@start}
\def\figure{\FIGNUM\? figure \?
   \immediate\write\figurewrite{\noexpand\refitem{\?.}}%
   \begingroup\obeyendofline\rw@start}
\def\Fig{\FIGNUM\? Fig. \?%
\immediate\write\figurewrite{\noexpand\refitem{\?.}}%
\begingroup\obeyendofline\rw@start}
\def\Figure{\FIGNUM\? Figure \?%
\immediate\write\figurewrite{\noexpand\refitem{\?.}}%
\begingroup\obeyendofline\rw@start}
\newcount\tablecount     \tablecount=0
\newif\iftableopen       \newwrite\tablewrite
\def\tabch@ck{\chardef\rw@write=\tablewrite \iftableopen\else
   \immediate\openout\tablewrite=tables.texauxil
   \tableopentrue\fi}
\def\TABNUM#1{\space@ver{}\tabch@ck \firstreflinetrue%
 \global\advance\tablecount by 1 \xdef#1{\the\tablecount}}
\def\TABLE#1{\TABNUM#1
   \immediate\write\tablewrite{\noexpand\refitem{#1.}}%
   \begingroup\obeyendofline\rw@start}
\def\Table{\TABNUM\? Table \?%
\immediate\write\tablewrite{\noexpand\refitem{\?.}}%
\begingroup\obeyendofline\rw@start}
\def\masterreset{\global\pagenumber=1 \global\chapternumber=0
   \ifnum\the\equanumber<0\else \global\equanumber=0\fi
   \global\sectionnumber=0
   \global\referencecount=0 \global\figurecount=0 \global\tablecount=0 }
\def\FRONTPAGE{\ifvoid255\else\vfill\penalty-2000\fi
      \masterreset\global\frontpagetrue
      \global\lastf@@t=0 \global\footsymbolcount=0}
\let\Frontpage=\FRONTPAGE
\def\paperstyle{\letterstylefalse\normalspace\papersize}

\def\papersize{\hsize=6.5in\vsize=9in\hoffset=0in\voffset=0in
               \skip\footins=\bigskipamount}

\paperstyle   
%
%
%
\newskip\frontpageskip
\newtoks\pubtype
\newtoks\Pubnum
\newtoks\pubnum
\newif\ifp@bblock  \p@bblocktrue
\def\PH@SR@V{\doubl@true \baselineskip=24.1pt plus 0.2pt minus 0.1pt
             \parskip= 3pt plus 2pt minus 1pt }
\def\PHYSREV{\paperstyle\PhysRevtrue\PH@SR@V}
\def\titlepage{\FRONTPAGE\paperstyle\ifPhysRev\PH@SR@V\fi
   \ifp@bblock\p@bblock\fi}
\def\nopubblock{\p@bblockfalse}
\def\endpage{\vfil\break}
\frontpageskip=1\medskipamount plus .5fil
\pubtype={\tensl Preliminary Version}
\def\title#1{\vskip\frontpageskip \titlestyle{#1} \vskip\headskip }
\def\author#1{\vskip\frontpageskip\titlestyle{\twelvecp #1}\nobreak}

\def\address#1{\par\kern 5pt\titlestyle{\twelvepoint\it #1}}
\def\andaddress{\par\kern 5pt \centerline{\sl and} \address}
\def\abstract{\vskip\frontpageskip\centerline{\fourteenrm ABSTRACT}
              \vskip\headskip }

\def\ie{\hbox{\it i.e.}}

\def\\{\relax\ifmmode\backslash\else$\backslash$\fi}
\def\globaleqnumbers{\relax\ifnum\the\equanumber<0%
\else\global\equanumber=-1\fi}

\def\journal#1&#2(#3){\unskip, \sl #1^\bf #2 \rm (19#3) }

\let\int=\intop         
\def\prop{\mathrel{{\mathchoice{\pr@p\scriptstyle}{\pr@p\scriptstyle}{
                \pr@p\scriptscriptstyle}{\pr@p\scriptscriptstyle} }}}
\def\pr@p#1{\setbox0=\hbox{$\cal #1 \char'103$}
   \hbox{$\cal #1 \char'117$\kern-.4\wd0\box0}}
\def\lsim{\mathrel{\mathpalette\@versim<}}
\def\gsim{\mathrel{\mathpalette\@versim>}}
\def\@versim#1#2{\lower0.2ex\vbox{\baselineskip\z@skip\lineskip\z@skip
  \lineskiplimit\z@\ialign{$\m@th#1\hfil##\hfil$\crcr#2\crcr\sim\crcr}}}
\let\sec@nt=\sec
\def\sec{\relax\ifmmode\let\n@xt=\sec@nt\else\let\n@xt\section\fi\n@xt}
\def\obsolete#1{\message{Macro \string #1 is obsolete.}}
\def\firstsec#1{\obsolete\firstsec \section{#1}}
\def\firstsubsec#1{\obsolete\firstsubsec \subsection{#1}}
\def\thispage#1{\obsolete\thispage \global\pagenumber=#1\frontpagefalse}
\def\thischapter#1{\obsolete\thischapter \global\chapternumber=#1}
\def\nextequation#1{\obsolete\nextequation \global\equanumber=#1
   \ifnum\the\equanumber>0 \global\advance\equanumber by 1 \fi}
\catcode`@=12 
\def\ifm#1{\relax\ifmmode#1\else$#1$\fi}

\def\gam{\ifm{\gamma}} \def\to{\ifm{\rightarrow}}

\def\pic{\ifm{\pi^+\pi^-}} \def\pio{\ifm{\pi^0\pi^0}}

\def\Kb{\ifm{\rlap{\kern.3em\raise1.9ex\hbox to.6em{\hrulefill}} K}}
\def\ab{\ifm{\sim}}  \def\x{\ifm{\times}}

\def\amp#1,#2,{\ifm{\langle#1|#2\rangle}}

\def\up#1{$^{#1}$}  
\def\etal{{\it et al.}}
\def\BR{{\rm BR}}

\def\ppc{\ifm{\pi^+\pi^-}}

\def\C{\ifm{C}}  \def\P{\ifm{P}}  \def\T{\ifm{T}}
\def\noc{\hglue.1pt\rlap{\raise .3mm\hbox{\kern.5mm\ifm{\backslash}\kern.7mm}}\C}
\def\nop{\hglue.1pt\rlap{\raise .3mm\hbox{\kern.5mm\ifm{\backslash}\kern.7mm}}\P}
\def\noT{\hglue.1pt\rlap{\raise .3mm\hbox{\kern.5mm\ifm{\backslash}\kern.7mm}}\T}
 
\def\pt#1,#2,{#1\x10\up{#2}}
\edef\epsfigRestoreAt{\catcode`@=\number\catcode`@\relax}%
\catcode`\@=11\relax
\ifx\undefined\@makeother                
\def\@makeother#1{\catcode`#1=12\relax}  
\fi                                      
\newcount\EPS@Height \newcount\EPS@Width \newcount\EPS@xscale
\newcount\EPS@yscale
\def\psfigdriver#1{%
  \bgroup\edef\next{\def\noexpand\tempa{#1}}%
    \uppercase\expandafter{\next}%
    \def\LN{DVITOLN03}%
    \def\DVItoPS{DVITOPS}%
    \def\DVIPS{DVIPS}%
    \def\emTeX{EMTEX}%
    \def\OzTeX{OZTEX}%
    \def\Textures{TEXTURES}%
    \global\chardef\fig@driver=0
    \ifx\tempa\LN
        \global\chardef\fig@driver=0\fi
    \ifx\tempa\DVItoPS
        \global\chardef\fig@driver=1\fi
    \ifx\tempa\DVIPS
        \global\chardef\fig@driver=2\fi
    \ifx\tempa\emTeX
        \global\chardef\fig@driver=3\fi
    \ifx\tempa\OzTeX
        \global\chardef\fig@driver=4\fi
    \ifx\tempa\Textures
        \global\chardef\fig@driver=5\fi
  \egroup
\def\psfig@start{}%
\def\psfig@end{}%
\def\epsfig@gofer{}%
\ifcase\fig@driver
\typeout{WARNING! ****
 no specials for LN03 psfig}%
\or 
\def\psfig@start{}%
\def\psfig@end{\special{dvitops: import \@p@sfilefinal \space
\@p@swidth sp \space \@p@sheight sp \space fill}%
\if@clip \typeout{Clipping not supported}\fi
\if@angle \typeout{Rotating not supported}\fi
}%
\let\epsfig@gofer\psfig@end
\or 
\def\psfig@start{\special{ps::[begin]  \@p@swidth \space \@p@sheight \space%
        \@p@sbbllx \space \@p@sbblly \space%
        \@p@sbburx \space \@p@sbbury \space%
        startTexFig \space }%
        \if@angle
                \special {ps:: \@p@sangle \space rotate \space}
        \fi
        \if@clip
                \if@verbose
                        \typeout{(clipped to BB) }%
                \fi
                \special{ps:: doclip \space }%
        \fi
        \special{ps: plotfile \@p@sfilefinal \space }%
        \special{ps::[end] endTexFig \space }%
}%
\def\psfig@end{}%
\def\epsfig@gofer{\if@clip
                        \if@verbose
                           \typeout{(clipped to BB)}%
                        \fi
                        \epsfclipon
                  \fi
                  \epsfsetgraph{\@p@sfilefinal}%
}%
\or 
\typeout{WARNING. You must have a .bb info file with the Bounding Box
  of the pcx file}%
\def\psfig@start{}%
\def\psfig@end{\typeout{pcx import of \@p@sfilefinal}%
\if@clip \typeout{Clipping not supported}\fi
\if@angle \typeout{Rotating not supported}\fi
\raisebox{\@p@srheight sp}{\special{em: graph \@p@sfilefinal}}}%
\def\epsfig@gofer{}%
\or 
\def\psfig@start{}%
\def\psfig@end{%
\EPS@Width\@p@swidth
\EPS@Height\@p@sheight
\divide\EPS@Width by 65781  
\divide\EPS@Height by 65781
\special{epsf=\@p@sfilefinal
\space
width=\the\EPS@Width
\space
height=\the\EPS@Height
}%
\if@clip \typeout{Clipping not supported}\fi
\if@angle \typeout{Rotating not supported}\fi
}%
\let\epsfig@gofer\psfig@end
\or 
\def\psfig@end{\if@clip
                        \if@verbose
                           \typeout{(clipped to BB)}%
                        \fi
                        \epsfclipon
                        \fi
\special{illustration \@p@sfilefinal\space scaled \the\EPS@xscale}%
}%
\def\psfig@start{}%
\let\epsfig\psfig
\else
\typeout{WARNING. *** unknown  driver - no psfig}%
\fi
}%
\newdimen\ps@dimcent
%
\ifx\undefined\fbox
\newdimen\fboxrule
\newdimen\fboxsep
\newdimen\ps@tempdima
\newbox\ps@tempboxa
\fboxsep = 0pt
\fboxrule = .4pt
\long\def\fbox#1{\leavevmode\setbox\ps@tempboxa\hbox{#1}\ps@tempdima\fboxrule
    \advance\ps@tempdima \fboxsep \advance\ps@tempdima \dp\ps@tempboxa
   \hbox{\lower \ps@tempdima\hbox
  {\vbox{\hrule height \fboxrule
          \hbox{\vrule width \fboxrule \hskip\fboxsep
          \vbox{\vskip\fboxsep \box\ps@tempboxa\vskip\fboxsep}\hskip
                 \fboxsep\vrule width \fboxrule}%
                 \hrule height \fboxrule}}}}%
\fi
\ifx\@ifundefined\undefined
\long\def\@ifundefined#1#2#3{\expandafter\ifx\csname
  #1\endcsname\relax#2\else#3\fi}%
\fi
\@ifundefined{typeout}%
{\gdef\typeout#1{\immediate\write\sixt@@n{#1}}}%
{\relax}%
\ifx\undefined\@latexerr
        \newlinechar`\^^J
        \def\@spaces{\space\space\space\space}%
        \def\@latexerr#1#2{%
        \edef\@tempc{#2}\expandafter\errhelp\expandafter{\@tempc}%
        \typeout{Error. \space see a manual for explanation.^^J
         \space\@spaces\@spaces\@spaces Type \space H <return> \space for
         immediate help.}\errmessage{#1}}%
\fi
%
\def\@nnil{\@nil}%
\def\@empty{}%
\def\@psdonoop#1\@@#2#3{}%
\def\@psdo#1:=#2\do#3{\edef\@psdotmp{#2}\ifx\@psdotmp\@empty \else
    \expandafter\@psdoloop#2,\@nil,\@nil\@@#1{#3}\fi}%
\def\@psdoloop#1,#2,#3\@@#4#5{\def#4{#1}\ifx #4\@nnil \else
       #5\def#4{#2}\ifx #4\@nnil \else#5\@ipsdoloop #3\@@#4{#5}\fi\fi}%
\def\@ipsdoloop#1,#2\@@#3#4{\def#3{#1}\ifx #3\@nnil
       \let\@nextwhile=\@psdonoop \else
      #4\relax\let\@nextwhile=\@ipsdoloop\fi\@nextwhile#2\@@#3{#4}}%
\def\@tpsdo#1:=#2\do#3{\xdef\@psdotmp{#2}\ifx\@psdotmp\@empty \else
    \@tpsdoloop#2\@nil\@nil\@@#1{#3}\fi}%
\def\@tpsdoloop#1#2\@@#3#4{\def#3{#1}\ifx #3\@nnil
       \let\@nextwhile=\@psdonoop \else
      #4\relax\let\@nextwhile=\@tpsdoloop\fi\@nextwhile#2\@@#3{#4}}%
%
%
\long\def\epsfaux#1#2:#3\\{\ifx#1\epsfpercent
   \def\testit{#2}\ifx\testit\epsfbblit
        \@atendfalse
        \epsf@atend #3 . \\%
        \if@atend
           \if@verbose
                \typeout{epsfig: found `(atend)'; continuing search}%
           \fi
        \else
                \epsfgrab #3 . . . \\%
                \epsffileokfalse\global\no@bbfalse
                \global\epsfbbfoundtrue
        \fi
   \fi\fi}%
\def\epsf@atendlit{(atend)}
\def\epsf@atend #1 #2 #3\\{%
   \def\epsf@tmp{#1}\ifx\epsf@tmp\empty
      \epsf@atend #2 #3 .\\\else
   \ifx\epsf@tmp\epsf@atendlit\@atendtrue\fi\fi}%

\chardef\trig@letter = 11
\chardef\other = 12
 
\newif\ifdebug 
\newif\ifc@mpute 
\newif\if@atend
\c@mputetrue 
 
\let\then = \relax
\def\r@dian{pt }%
\let\r@dians = \r@dian
\let\dimensionless@nit = \r@dian
\let\dimensionless@nits = \dimensionless@nit
\def\internal@nit{sp }%
\let\internal@nits = \internal@nit
\newif\ifstillc@nverging
\def \Mess@ge #1{\ifdebug \then \message {#1} \fi}%
 
{ 
        \catcode `\@ = \trig@letter
        \gdef \nodimen {\expandafter \n@dimen \the \dimen}%
        \gdef \term #1 #2 #3%
               {\edef \t@ {\the #1}
                \edef \t@@ {\expandafter \n@dimen \the #2\r@dian}%
                \t@rm {\t@} {\t@@} {#3}%
               }%
        \gdef \t@rm #1 #2 #3%
               {{%
                \count 0 = 0
                \dimen 0 = 1 \dimensionless@nit
                \dimen 2 = #2\relax
                \Mess@ge {Calculating term #1 of \nodimen 2}%
                \loop
                \ifnum  \count 0 < #1
                \then   \advance \count 0 by 1
                        \Mess@ge {Iteration \the \count 0 \space}%
                        \Multiply \dimen 0 by {\dimen 2}%
                        \Mess@ge {After multiplication, term = \nodimen 0}%
                        \Divide \dimen 0 by {\count 0}%
                        \Mess@ge {After division, term = \nodimen 0}%
                \repeat
                \Mess@ge {Final value for term #1 of
                                \nodimen 2 \space is \nodimen 0}%
                \xdef \Term {#3 = \nodimen 0 \r@dians}%
                \aftergroup \Term
               }}%
        \catcode `\p = \other
        \catcode `\t = \other
        \gdef \n@dimen #1pt{#1} 
}%
 
\def \Divide #1by #2{\divide #1 by #2} 
 
\def \Multiply #1by #2
       {{
        \count 0 = #1\relax
        \count 2 = #2\relax
        \count 4 = 65536
        \Mess@ge {Before scaling, count 0 = \the \count 0 \space and
                        count 2 = \the \count 2}%
        \ifnum  \count 0 > 32767 
        \then   \divide \count 0 by 4
                \divide \count 4 by 4
        \else   \ifnum  \count 0 < -32767
                \then   \divide \count 0 by 4
                        \divide \count 4 by 4
                \else
                \fi
        \fi
        \ifnum  \count 2 > 32767 
        \then   \divide \count 2 by 4
                \divide \count 4 by 4
        \else   \ifnum  \count 2 < -32767
                \then   \divide \count 2 by 4
                        \divide \count 4 by 4
                \else
                \fi
        \fi
        \multiply \count 0 by \count 2
        \divide \count 0 by \count 4
        \xdef \product {#1 = \the \count 0 \internal@nits}%
        \aftergroup \product
       }}%
 
\def\r@duce{\ifdim\dimen0 > 90\r@dian \then   
                \multiply\dimen0 by -1
                \advance\dimen0 by 180\r@dian
                \r@duce
            \else \ifdim\dimen0 < -90\r@dian \then  
                \advance\dimen0 by 360\r@dian
                \r@duce
                \fi
            \fi}%
 
\def\Sine#1%
       {{%
        \dimen 0 = #1 \r@dian
        \r@duce
        \ifdim\dimen0 = -90\r@dian \then
           \dimen4 = -1\r@dian
           \c@mputefalse
        \fi
        \ifdim\dimen0 = 90\r@dian \then
           \dimen4 = 1\r@dian
           \c@mputefalse
        \fi
        \ifdim\dimen0 = 0\r@dian \then
           \dimen4 = 0\r@dian
           \c@mputefalse
        \fi
        \ifc@mpute \then
                \divide\dimen0 by 180
                \dimen0=3.141592654\dimen0
                \dimen 2 = 3.1415926535897963\r@dian 
                \divide\dimen 2 by 2 
                \Mess@ge {Sin: calculating Sin of \nodimen 0}%
                \count 0 = 1 
                \dimen 2 = 1 \r@dian 
                \dimen 4 = 0 \r@dian 
                \loop
                        \ifnum  \dimen 2 = 0 
                        \then   \stillc@nvergingfalse
                        \else   \stillc@nvergingtrue
                        \fi
                        \ifstillc@nverging 
                        \then   \term {\count 0} {\dimen 0} {\dimen 2}%
                                \advance \count 0 by 2
                                \count 2 = \count 0
                                \divide \count 2 by 2
                                \ifodd  \count 2 
                                \then   \advance \dimen 4 by \dimen 2
                                \else   \advance \dimen 4 by -\dimen 2
                                \fi
                \repeat
        \fi
                        \xdef \sine {\nodimen 4}%
       }}%
 
\def\Cosine#1{\ifx\sine\UnDefined\edef\Savesine{\relax}\else
                             \edef\Savesine{\sine}\fi
        {\dimen0=#1\r@dian\multiply\dimen0 by -1
         \advance\dimen0 by 90\r@dian
         \Sine{\nodimen 0}%
         \xdef\cosine{\sine}%
         \xdef\sine{\Savesine}}}
%
\def\psdraft{\def\@psdraft{0}}%
\def\psfull{\def\@psdraft{1}}%
\psfull
\newif\if@compress
\def\pscompress{\@compresstrue}
\def\psnocompress{\@compressfalse}
\@compressfalse
\newif\if@scalefirst
\def\psscalefirst{\@scalefirsttrue}%
\def\psrotatefirst{\@scalefirstfalse}%
\psrotatefirst
\newif\if@draftbox
\def\psnodraftbox{\@draftboxfalse}%
\@draftboxtrue
\newif\if@noisy
\@noisyfalse
\newif\ifno@bb
\newif\if@bbllx
\newif\if@bblly
\newif\if@bburx
\newif\if@bbury
\newif\if@height
\newif\if@width
\newif\if@rheight
\newif\if@rwidth
\newif\if@angle
\newif\if@clip
\newif\if@verbose
\newif\if@prologfile
\def\@p@@sprolog#1{\@prologfiletrue\def\@prologfileval{#1}}%
\def\@p@@sclip#1{\@cliptrue}%
\newif\ifepsfig@dos  
\def\epsfigdos{\epsfig@dostrue}%
\epsfig@dosfalse
\newif\ifuse@psfig
\def\ParseName#1{\expandafter\@Parse#1}%
\def\@Parse#1.#2:{\gdef\BaseName{#1}\gdef\FileType{#2}}%

\def\@p@@sfile#1{%
  \ifepsfig@dos
     \ParseName{#1:}%
  \else
     \gdef\BaseName{#1}\gdef\FileType{}%
  \fi
  \def\@p@sfile{NO FILE: #1}%
  \def\@p@sfilefinal{NO FILE: #1}%
  \openin1=#1
  \ifeof1\closein1\openin1=\BaseName.bb
    \ifeof1\closein1
      \if@bbllx                 
        \if@bblly\if@bburx\if@bbury
          \def\@p@sfile{#1}%
          \def\@p@sfilefinal{#1}%
        \fi\fi\fi
      \else                     
        \@latexerr{ERROR. PostScript file #1 not found}\@whattodo
        \@p@@sbbllx{100bp}%
        \@p@@sbblly{100bp}%
        \@p@@sbburx{200bp}%
        \@p@@sbbury{200bp}%
        \psdraft
      \fi
    \else                       
      \closein1%
      \edef\@p@sfile{\BaseName.bb}%
      \typeout{using BB from \@p@sfile}%
      \ifnum\fig@driver=3
        \edef\@p@sfilefinal{\BaseName.pcx}%
      \else
        \ifepsfig@dos
          \edef\@p@sfilefinal{"`gunzip -c `texfind \BaseName.{z,Z,gz}`"}%
        \else
          \edef\@p@sfilefinal{"`epsfig \if@compress-c \fi#1"}%
        \fi
      \fi
    \fi
  \else\closein1                
    \edef\@p@sfile{#1}%
    \if@compress  
      \edef\@p@sfilefinal{"`epsfig -c #1"}%
    \else
      \edef\@p@sfilefinal{#1}%
    \fi
  \fi%
}

\let\@p@@sfigure\@p@@sfile
\def\@p@@sbbllx#1{%
                                            \@bbllxtrue
                \ps@dimcent=#1
                \edef\@p@sbbllx{\number\ps@dimcent}%
                \divide\ps@dimcent by65536
                \global\edef\epsfllx{\number\ps@dimcent}%
}%
\def\@p@@sbblly#1{%
                \@bbllytrue
                \ps@dimcent=#1
                \edef\@p@sbblly{\number\ps@dimcent}%
                \divide\ps@dimcent by65536
                \global\edef\epsflly{\number\ps@dimcent}%
}%
\def\@p@@sbburx#1{%
                \@bburxtrue
                \ps@dimcent=#1
                \edef\@p@sbburx{\number\ps@dimcent}%
                \divide\ps@dimcent by65536
                \global\edef\epsfurx{\number\ps@dimcent}%
}%
\def\@p@@sbbury#1{%
                \@bburytrue
                \ps@dimcent=#1
                \edef\@p@sbbury{\number\ps@dimcent}%
                \divide\ps@dimcent by65536
                \global\edef\epsfury{\number\ps@dimcent}%
}%
\def\@p@@sheight#1{%
                \@heighttrue
                \global\epsfysize=#1
                \ps@dimcent=#1
                \edef\@p@sheight{\number\ps@dimcent}%
}%
\def\@p@@swidth#1{%
                \@widthtrue
                \global\epsfxsize=#1
                \ps@dimcent=#1
                \edef\@p@swidth{\number\ps@dimcent}%
}%
\def\@p@@srheight#1{%
                \@rheighttrue\use@psfigtrue
                \ps@dimcent=#1
                \edef\@p@srheight{\number\ps@dimcent}%
}%
\def\@p@@srwidth#1{%
                \@rwidthtrue\use@psfigtrue
                \ps@dimcent=#1
                \edef\@p@srwidth{\number\ps@dimcent}%
}%
\def\@p@@sangle#1{%
                \use@psfigtrue
                \@angletrue
                \edef\@p@sangle{#1}%
}%
\def\@p@@ssilent#1{%
                \@verbosefalse
}%
\def\@p@@snoisy#1{%
                \@verbosetrue
}%
\def\@cs@name#1{\csname #1\endcsname}%
\def\@setparms#1=#2,{\@cs@name{@p@@s#1}{#2}}%
%
%
\def\ps@init@parms{%
                \@bbllxfalse \@bbllyfalse
                \@bburxfalse \@bburyfalse
                \@heightfalse \@widthfalse
                \@rheightfalse \@rwidthfalse
                \def\@p@sbbllx{}\def\@p@sbblly{}%
                \def\@p@sbburx{}\def\@p@sbbury{}%
                \def\@p@sheight{}\def\@p@swidth{}%
                \def\@p@srheight{}\def\@p@srwidth{}%
                \def\@p@sangle{0}%
                \def\@p@sfile{}%
                \use@psfigfalse
                \@prologfilefalse
                \def\@sc{}%
                \if@noisy
                        \@verbosetrue
                \else
                        \@verbosefalse
                \fi
                \@clipfalse
}%
%
%
\def\parse@ps@parms#1{%
                \@psdo\@psfiga:=#1\do
                   {\expandafter\@setparms\@psfiga,}%
\if@prologfile
\fi
}%
%
%
\def\bb@missing{%
        \if@verbose
            \typeout{psfig: searching \@p@sfile \space  for bounding box}%
        \fi
        \epsfgetbb{\@p@sfile}%
        \ifepsfbbfound
            \ps@dimcent=\epsfllx bp\edef\@p@sbbllx{\number\ps@dimcent}%
            \ps@dimcent=\epsflly bp\edef\@p@sbblly{\number\ps@dimcent}%
            \ps@dimcent=\epsfurx bp\edef\@p@sbburx{\number\ps@dimcent}%
            \ps@dimcent=\epsfury bp\edef\@p@sbbury{\number\ps@dimcent}%
        \else
            \epsfbbfoundfalse
        \fi
}
%
\newdimen\p@intvaluex
\newdimen\p@intvaluey
\def\rotate@#1#2{{\dimen0=#1 sp\dimen1=#2 sp
                  \global\p@intvaluex=\cosine\dimen0
                  \dimen3=\sine\dimen1
                  \global\advance\p@intvaluex by -\dimen3
                  \global\p@intvaluey=\sine\dimen0
                  \dimen3=\cosine\dimen1
                  \global\advance\p@intvaluey by \dimen3
                  }}%
\def\compute@bb{%
                \epsfbbfoundfalse
                \if@bbllx\epsfbbfoundtrue\fi
                \if@bblly\epsfbbfoundtrue\fi
                \if@bburx\epsfbbfoundtrue\fi
                \if@bbury\epsfbbfoundtrue\fi
                \ifepsfbbfound\else\bb@missing\fi
                \ifepsfbbfound\else
                \@latexerr{ERROR. cannot locate BoundingBox}\@whattodobb
                        \@p@@sbbllx{100bp}%
                        \@p@@sbblly{100bp}%
                        \@p@@sbburx{200bp}%
                        \@p@@sbbury{200bp}%
                        \no@bbtrue
                        \psdraft
                \fi
                %
%
                \count203=\@p@sbburx
                \count204=\@p@sbbury
                \advance\count203 by -\@p@sbbllx
                \advance\count204 by -\@p@sbblly
                \edef\ps@bbw{\number\count203}%
                \edef\ps@bbh{\number\count204}%
                 \edef\@bbw{\number\count203}%
                \edef\@bbh{\number\count204}%
               \if@angle
                        \Sine{\@p@sangle}\Cosine{\@p@sangle}%
 
{\ps@dimcent=\maxdimen\xdef\r@p@sbbllx{\number\ps@dimcent}%
 
\xdef\r@p@sbblly{\number\ps@dimcent}%
 
\xdef\r@p@sbburx{-\number\ps@dimcent}%
 
\xdef\r@p@sbbury{-\number\ps@dimcent}}%
%
                        \def\minmaxtest{%
                           \ifnum\number\p@intvaluex<\r@p@sbbllx
                              \xdef\r@p@sbbllx{\number\p@intvaluex}\fi
                           \ifnum\number\p@intvaluex>\r@p@sbburx
                              \xdef\r@p@sbburx{\number\p@intvaluex}\fi
                           \ifnum\number\p@intvaluey<\r@p@sbblly
                              \xdef\r@p@sbblly{\number\p@intvaluey}\fi
                           \ifnum\number\p@intvaluey>\r@p@sbbury
                              \xdef\r@p@sbbury{\number\p@intvaluey}\fi
                           }%
                        \rotate@{\@p@sbbllx}{\@p@sbblly}%
                        \minmaxtest
                        \rotate@{\@p@sbbllx}{\@p@sbbury}%
                        \minmaxtest
                        \rotate@{\@p@sbburx}{\@p@sbblly}%
                        \minmaxtest
                        \rotate@{\@p@sbburx}{\@p@sbbury}%
                        \minmaxtest
 
\edef\@p@sbbllx{\r@p@sbbllx}\edef\@p@sbblly{\r@p@sbblly}%
 
\edef\@p@sbburx{\r@p@sbburx}\edef\@p@sbbury{\r@p@sbbury}%
                \fi
                \count203=\@p@sbburx
                \count204=\@p@sbbury
                \advance\count203 by -\@p@sbbllx
                \advance\count204 by -\@p@sbblly
                \edef\@bbw{\number\count203}%
                \edef\@bbh{\number\count204}%
}%
%
%
\def\in@hundreds#1#2#3{\count240=#2 \count241=#3
                     \count100=\count240        
                     \divide\count100 by \count241
                     \count101=\count100
                     \multiply\count101 by \count241
                     \advance\count240 by -\count101
                     \multiply\count240 by 10
                     \count101=\count240        
                     \divide\count101 by \count241
                     \count102=\count101
                     \multiply\count102 by \count241
                     \advance\count240 by -\count102
                     \multiply\count240 by 10
                     \count102=\count240        
                     \divide\count102 by \count241
                     \count200=#1\count205=0
                     \count201=\count200
                        \multiply\count201 by \count100
                        \advance\count205 by \count201
                     \count201=\count200
                        \divide\count201 by 10
                        \multiply\count201 by \count101
                        \advance\count205 by \count201
                     \count201=\count200
                        \divide\count201 by 100
                        \multiply\count201 by \count102
                        \advance\count205 by \count201
                     \edef\@result{\number\count205}%
}%
\def\compute@wfromh{%
                \in@hundreds{\@p@sheight}{\@bbw}{\@bbh}%
                \edef\@p@swidth{\@result}%
}%
\def\compute@hfromw{%
                \in@hundreds{\@p@swidth}{\@bbh}{\@bbw}%
                \edef\@p@sheight{\@result}%
}%
\def\compute@handw{%
                \if@height
                        \if@width
                        \else
                                \compute@wfromh
                        \fi
                \else
                        \if@width
                                \compute@hfromw
                        \else
                                \edef\@p@sheight{\@bbh}%
                                \edef\@p@swidth{\@bbw}%
                        \fi
                \fi
}%
\def\compute@resv{%
                \if@rheight \else \edef\@p@srheight{\@p@sheight} \fi
                \if@rwidth \else \edef\@p@srwidth{\@p@swidth} \fi
}%
%
\def\compute@sizes{%
        \if@scalefirst\if@angle
        \if@width
           \in@hundreds{\@p@swidth}{\@bbw}{\ps@bbw}%
           \edef\@p@swidth{\@result}%
        \fi
        \if@height
           \in@hundreds{\@p@sheight}{\@bbh}{\ps@bbh}%
           \edef\@p@sheight{\@result}%
        \fi
        \fi\fi
        \compute@handw
        \compute@resv
                                                   \EPS@Width=\@bbw  
                                                                                                                                \divide\EPS@Width by 1000
                                                                                                 \EPS@xscale=\@p@swidth \divide \EPS@xscale by \EPS@Width
                                                   \EPS@Height=\@bbh  
                                                                                                                                \divide\EPS@Height by 1000
                                                                                                 \EPS@yscale=\@p@sheight \divide \EPS@yscale by\EPS@Height
  \ifnum\EPS@xscale>\EPS@yscale\EPS@xscale=\EPS@yscale\fi
}
%

\long\def\graphic@verb#1{\def\next{#1}%
  {\expandafter\graphic@strip\meaning\next}}
\def\graphic@strip#1>{}
\def\graphic@zapspace#1{%
  #1\ifx\graphic@zapspace#1\graphic@zapspace%
  \else\expandafter\graphic@zapspace%
  \fi}
\def\psfig#1{%
\edef\@tempa{\graphic@zapspace#1{}}%
\ifvmode\leavevmode\fi\vbox {%
        \ps@init@parms
        \parse@ps@parms{\@tempa}%
        \ifnum\@psdraft=1
                \typeout{[\@p@sfilefinal]}%
                \if@verbose
                        \typeout{epsfig: using PSFIG macros}%
                \fi
                \psfig@method
        \else
                \epsfig@draft
        \fi
}
}%
\def\graphic@zapspace#1{%
  #1\ifx\graphic@zapspace#1\graphic@zapspace%
  \else\expandafter\graphic@zapspace%
  \fi}
\def\epsfig#1{%
\edef\@tempa{\graphic@zapspace#1{}}%
\ifvmode\leavevmode\fi\vbox {%
        \ps@init@parms
        \parse@ps@parms{\@tempa}%
        \ifnum\@psdraft=1
          \if@angle\use@psfigtrue\fi
          {\ifnum\fig@driver=1\global\use@psfigtrue\fi}%
          {\ifnum\fig@driver=3\global\use@psfigtrue\fi}%
          {\ifnum\fig@driver=4\global\use@psfigtrue\fi}%
          {\ifnum\fig@driver=5\global\use@psfigtrue\fi}%
                \ifuse@psfig
                        \if@verbose
                                \typeout{epsfig: using PSFIG macros}%
                        \fi
                        \psfig@method
                \else
                        \if@verbose
                                \typeout{epsfig: using EPSF macros}%
                        \fi
                        \epsf@method
                \fi
        \else
                \epsfig@draft
        \fi
}%
}%

\def\epsf@method{%
        \epsfbbfoundfalse
        \if@bbllx\epsfbbfoundtrue\fi
        \if@bblly\epsfbbfoundtrue\fi
        \if@bburx\epsfbbfoundtrue\fi
        \if@bbury\epsfbbfoundtrue\fi
        \ifepsfbbfound\else\epsfgetbb{\@p@sfile}\fi
        \ifepsfbbfound
           \typeout{<\@p@sfilefinal>}%
           \epsfig@gofer
        \else
          \@latexerr{ERROR - Cannot locate BoundingBox}\@whattodobb
          \@p@@sbbllx{100bp}%
          \@p@@sbblly{100bp}%
          \@p@@sbburx{200bp}%
          \@p@@sbbury{200bp}%
                \count203=\@p@sbburx
                \count204=\@p@sbbury
                \advance\count203 by -\@p@sbbllx
                \advance\count204 by -\@p@sbblly
                \edef\@bbw{\number\count203}%
                \edef\@bbh{\number\count204}%
          \compute@sizes
          \epsfig@@draft
       \fi
}%
\def\psfig@method{%
        \compute@bb
        \ifepsfbbfound
          \compute@sizes
          \psfig@start
          \vbox to \@p@srheight sp{\hbox to \@p@srwidth 
            sp{\hss}\vss\psfig@end}%
        \else
           \epsfig@draft
        \fi
}%
%
\def\epsfig@draft{\compute@bb\compute@sizes\epsfig@@draft}%
\def\epsfig@@draft{%
\typeout{<(draft only) \@p@sfilefinal>}%
\if@draftbox
        \hbox{{\fboxsep0pt\fbox{\vbox to \@p@srheight sp{%
        \vss\hbox to \@p@srwidth sp{ \hss 
           \expandafter\Literally\@p@sfilefinal\@nil
                          \hss }\vss
        }}}}%
\else
        \vbox to \@p@srheight sp{%
        \vss\hbox to \@p@srwidth sp{\hss}\vss}%
\fi
}%
\def\Literally#1\@nil{{\tt\graphic@verb{#1}}}
\psfigdriver{dvips}%
\epsfigRestoreAt
\newread\epsffilein    
\newif\ifepsffileok    
\newif\ifepsfbbfound   
\newif\ifepsfverbose   
\newdimen\epsfxsize    
\newdimen\epsfysize    
\newdimen\epsftsize    
\newdimen\epsfrsize    
\newdimen\epsftmp      
\newdimen\pspoints     
\pspoints=1bp          
\epsfxsize=0pt         
\epsfysize=0pt         
\def\epsfbox#1{\global\def\epsfllx{72}\global\def\epsflly{72}%
   \global\def\epsfurx{540}\global\def\epsfury{720}%
   \def\lbracket{[}\def\testit{#1}\ifx\testit\lbracket
   \let\next=\epsfgetlitbb\else\let\next=\epsfnormal\fi\next{#1}}%
\def\epsfgetlitbb#1#2 #3 #4 #5]#6{\epsfgrab #2 #3 #4 #5 .\\%
   \epsfsetgraph{#6}}%
\def\epsfnormal#1{\epsfgetbb{#1}\epsfsetgraph{#1}}%
\def\epsfgetbb#1{%
\openin\epsffilein=#1
\ifeof\epsffilein\errmessage{I couldn't open #1, will ignore it}\else
   {\epsffileoktrue \chardef\other=12
    \def\do##1{\catcode`##1=\other}\dospecials \catcode`\ =10
    \loop
       \read\epsffilein to \epsffileline
       \ifeof\epsffilein\epsffileokfalse\else
          \expandafter\epsfaux\epsffileline:. \\%
       \fi
   \ifepsffileok\repeat
   \ifepsfbbfound\else
    \ifepsfverbose\message{No bounding box comment in #1; using defaults}\fi\fi
   }\closein\epsffilein\fi}%
\def\epsfclipstring{}
\def\epsfclipon{\def\epsfclipstring{ clip}}%
\def\epsfsetgraph#1{%
   \epsfrsize=\epsfury\pspoints
   \advance\epsfrsize by-\epsflly\pspoints
   \epsftsize=\epsfurx\pspoints
   \advance\epsftsize by-\epsfllx\pspoints
   \epsfxsize\epsfsize\epsftsize\epsfrsize
   \ifnum\epsfxsize=0 \ifnum\epsfysize=0
      \epsfxsize=\epsftsize \epsfysize=\epsfrsize
      \epsfrsize=0pt
     \else\epsftmp=\epsftsize \divide\epsftmp\epsfrsize
       \epsfxsize=\epsfysize \multiply\epsfxsize\epsftmp
       \multiply\epsftmp\epsfrsize \advance\epsftsize-\epsftmp
       \epsftmp=\epsfysize
       \loop \advance\epsftsize\epsftsize \divide\epsftmp 2
       \ifnum\epsftmp>0
          \ifnum\epsftsize<\epsfrsize\else
             \advance\epsftsize-\epsfrsize \advance\epsfxsize\epsftmp \fi
       \repeat
       \epsfrsize=0pt
     \fi
   \else \ifnum\epsfysize=0
     \epsftmp=\epsfrsize \divide\epsftmp\epsftsize
     \epsfysize=\epsfxsize \multiply\epsfysize\epsftmp   
     \multiply\epsftmp\epsftsize \advance\epsfrsize-\epsftmp
     \epsftmp=\epsfxsize
     \loop \advance\epsfrsize\epsfrsize \divide\epsftmp 2
     \ifnum\epsftmp>0
        \ifnum\epsfrsize<\epsftsize\else
           \advance\epsfrsize-\epsftsize \advance\epsfysize\epsftmp \fi
     \repeat
     \epsfrsize=0pt
    \else
     \epsfrsize=\epsfysize
    \fi
   \fi
   \ifepsfverbose\message{#1: width=\the\epsfxsize, height=\the\epsfysize}\fi
   \epsftmp=10\epsfxsize \divide\epsftmp\pspoints
   \vbox to\epsfysize{\vfil\hbox to\epsfxsize{%
      \ifnum\epsfrsize=0\relax
        \includegraphics{#1}%
      \else
        \epsfrsize=10\epsfysize \divide\epsfrsize\pspoints
        \includegraphics{#1}%
      \fi
      \hfil}}%
\global\epsfxsize=0pt\global\epsfysize=0pt}%
{\catcode`\%=12 \global\let\epsfpercent=
\long\def\epsfaux#1#2:#3\\{\ifx#1\epsfpercent
   \def\testit{#2}\ifx\testit\epsfbblit
      \epsfgrab #3 . . . \\%
      \epsffileokfalse
      \global\epsfbbfoundtrue
   \fi\else\ifx#1\par\else\epsffileokfalse\fi\fi}%
\def\epsfempty{}%
\def\epsfgrab #1 #2 #3 #4 #5\\{%
\global\def\epsfllx{#1}\ifx\epsfllx\epsfempty
      \epsfgrab #2 #3 #4 #5 .\\\else
   \global\def\epsflly{#2}%
   \global\def\epsfurx{#3}\global\def\epsfury{#4}\fi}%
\def\epsfsize#1#2{\epsfxsize}
%


\let\cl=\centerline

\def\Y{\ifm{\Upsilon}}
\def\Up#1{\ifm{\Upsilon(#1 S)}}

\def\Ypp{\ifm{\Upsilon''}}
\def\X{\ifm{\chi _b}}
\def\Xp{\ifm{\chi _b'}}
\def\muu{\ifm{\mu^+\mu^-}}
\def\bbar{\ifm{b\overline b}} 

\def\B{\ifm{B}} \def\BS{\ifm{B_s}} \def\Bs{\ifm{B^*}} 
\def\Ups{\ifm{\Upsilon}}  \def\Uvs{\ifm{\Upsilon(5S)}}
\def\chib{\ifm{\chi_b}} 

\def\xb#1){\ifm{\chi_b#1)}}  \def\PM{\ifm{\pm}}
\def\undert#1{$\underline{\hbox{#1}}$}

\FIG\uplevel
\FIG\cusbzero
\FIG\cusb
\FIG\rmeter
\FIG\bgo
\FIG\bstarb
\FIG\eone
\FIG\gamspec
\FIG\FineS
\FIG\Fscal
\FIG\dipion

\def\refmark#1{(#1)}
\def\refend{ \refmark{\number\referencecount}\kern-.8ex}
\def\refsend{\refmark{\count255=\referencecount
   \advance\count255 by-\lastrefsbegincount
   \ifcase\count255 \number\referencecount
   \or \number\lastrefsbegincount,\number\referencecount
   \else \number\lastrefsbegincount-\number\referencecount \fi}}

\overfullrule 0pt
\singlespace

\hsize=150 mm \hoffset=.55cm
\vsize=219 mm

\Frontpage

\vbox to 4.5cm{
\vglue1cm
\line{\hfill\undertext{LNF-97/031 (P)}}
\line{\hfill 2 September 1997}
\vfill}

\font\twpointb=cmbx10 at 20 pt

\vglue5mm
\cl{\twpointb Hidden and Open Beauty in CUSB}
\vglue10mm
\cl{\fourteenpoint Juliet Lee-Franzini}
\vglue 10mm

\baselineskip14pt
\cl{\it Laboratori di Frascati dell'INFN}
\cl{\it CP-13, Via Enrico Fermi 40, I-00044, Frascati Roma and}
\cl{\it Physics Department, SUNY at Stony Brook, N.Y. 11794, U.S.A.}

\vfill

\vbox{\parshape=2 6mm 136.5mm 6mm 136.5mm
\baselineskip13pt
\noindent{\bf Abstract.} We present a brief history and a summary of the physics
results of CUSB at CESR.}
\vglue1cm
\eject

\par\hskip0pt

\baselineskip13pt
\cl{\fourteenpoint\bf HISTORY}
\vglue3mm

\REFS\jlfsur{Lee-Franzini, J. ``Quark Spectroscopy with a new flavor", 
Survey in High Energy Physics {\bf 2}, 3 (1981).}
\REFSCON\pjpr{Franzini, P. and Lee-Franzini, J., ``Upsilon Physics at CESR", 
Physics Report {\bf81C},  3 (1982).}
\REFSCON\pjarn{Franzini, P. and Lee-Franzini, J., ``Upsilon Resonances", 
Ann. Rev. Nuc. Part. Sci., {\bf 33}, 1 (1983).}
\REFSCON\jlfmix{Lee-Franzini, J., Ono, S., Tornqvist, N. A. and Sanda, A.I., 
``Where Are the $B\overline B$ Mixing Effects Observable in the
\Y's Region?", 
Phys. Rev. Lett. {\bf 55} 2938 (1985).}
\REFSCON\pjips{Franzini, P. and Lee-Franzini, J., ``10 Years of 
\Y\ Spectroscopy", {\it Present and 
Future of Collider Physics}, ed. C. Bacci \etal, Italian Phys. Soc.
Conf. Proc., Editrice Compositori, Bologna, 255 (1991).} 
\REFSCON\pafs{Franzini, Paula J., ``Fine Structure of the 
P States in Quarkonium Fine Structure 
and the nature of the Spin-dependent Potential",
Phys. Lett. {\bf B296} 199 (1992).}
\REFSCON\pasv{Franzini, Paula J., ``A Pseudoscalar Field in the 
q$\bar q$ Interaction in Heavy Quarkonium?", Phys. Lett. {\bf B296} 195 (1992).}
\REFSCON\pajhad{Lee-Franzini, J. and Franzini, Paula J.,  
``Hadronic Width of the \X\ States'',
{\it Proc. of the Third Workshop on the Tau-Charm Factory},
Marbella, Spain, 457 (1993).}
\REFSCON\wuthesis{Wu, Q. W., ``Study of \ppc\ transitions from Upsilons",
thesis, Columbia University, Nevis Labs. (1993).}
\baselineskip13pt
\noindent
Since the discovery of the \Y\ system at FNAL, as told at the begining 
of this volume, the CUSB (Columbia
University-Stony Brook) collaboration at CESR (Cornell Electron Storage
Ring) has played a pivotal role in the study of the upsilon system,
potential models, QCD, and weak interactions of the b quark \refsend.  
In retrospect, things happened very fast. CUSB's formal proposal 
was submitted to Cornell on December 2nd, 1977
and was approved by February 15th, 1978! 

\vglue3mm
\vbox{\centerline{\epsfig{file=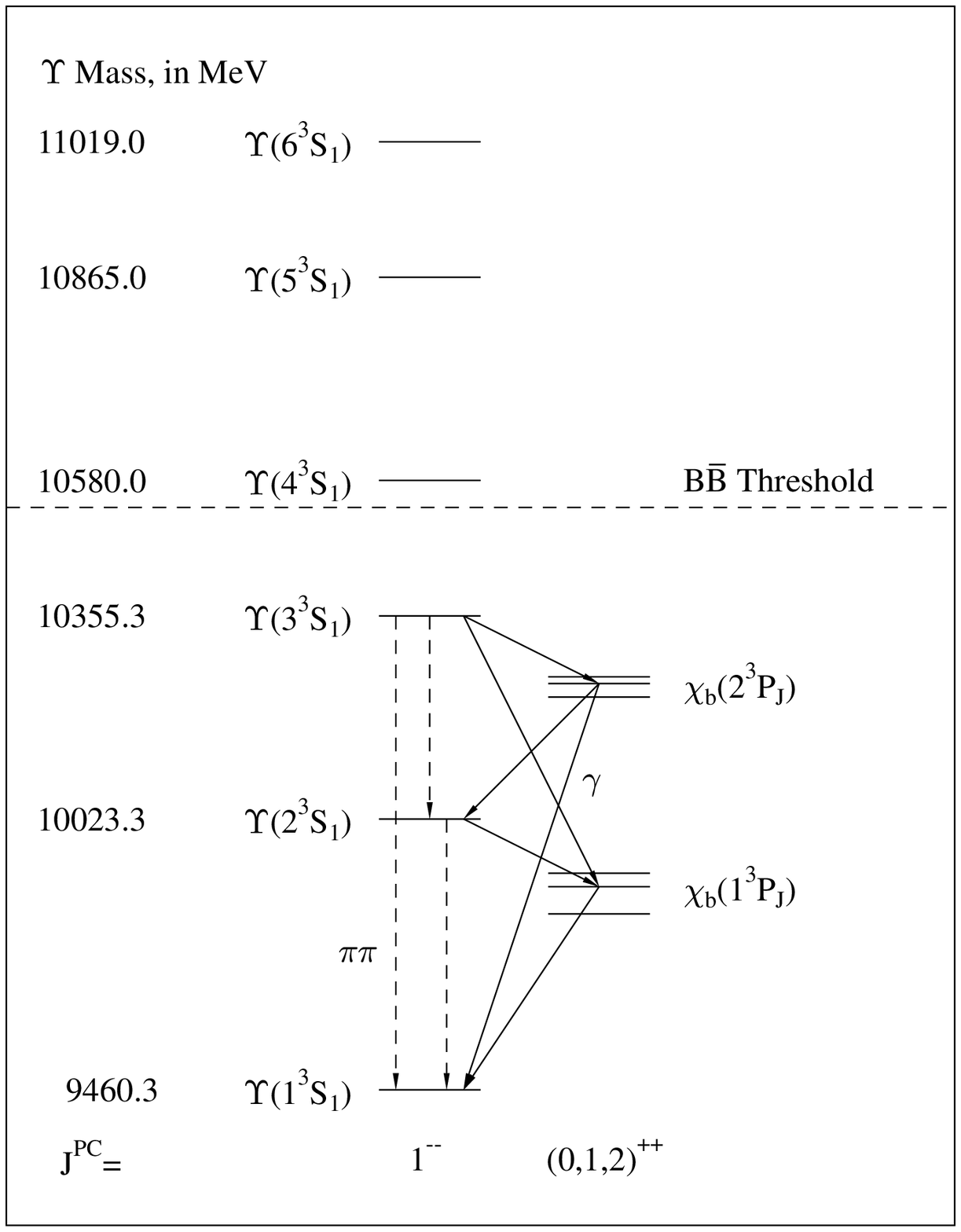,height=14cm}}}  
\vglue1mm
\vbox{\parshape=1 6mm 136.5mm 
\tenpoint\baselineskip10pt
\noindent{\bf Figure \uplevel.} \Y's Level Diagram, CUSB observed 
and studied all level and transitions between levels indicated above.
Open beauty was discovered in the decay \Up4\to high energy electrons. 
The \Bs's were found in the continuum above the \Up4.}
\vglue3mm

\noindent
On August 16th, 1979 we paid our first visit to the North Area (NA)
underground {\it cul de sac} which became the home for our electronics.
In May 1979 we installed the first NaI qradrant at the NA interaction 
region. In September of 1979 a caravan of two trailers containing electronics, 
escorted by a truck and passenger cars driven by and containing a dozen
physicists, arrived to the Northern edge of the Cornell Alumni 
Atheletic Fields. The trailers functioned as the CUSB data acquisition 
center, until May 1991 when they were honorably retired, their outer
skins showing various indentations resulting from seasonal sport projectiles.
On October 18th, 1979, with half of the NAI detector, 8.3 radiation length
(X$_0$) deep, centered upon the NA interaction area, we observed 
the first Bhabha ever seen at CESR.

\vfill
\vbox{\centerline{\epsfig{file=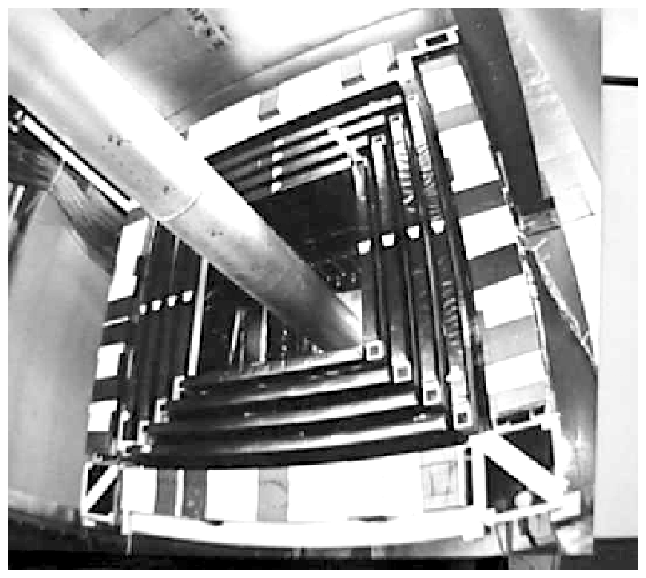,height=9.5cm}}}  
\vglue2mm
\vbox{\parshape=1 6mm 136.5mm 
\tenpoint\baselineskip10pt
\noindent{\bf Figure \cusbzero.} In this form, \ie\ just one half of the 
NaI crystal array, CUSB observed the three \Y's below threshold and on 
December 1979 had evidence for the \Up4, the first level above 
threshold.}
\vglue2mm\noindent
From then until January 1984,
in between collecting 950,000 events, we completed CUSB-I.
The CUSB-I detector comprises of 320 rectangular NaI crystals in a 
square geometry, surrounded by an array of Lead glass cubes 
($15\x15\x17.5$ cm$^3$, 7.7 X$_0$) for shower leakage containment.
These, in turn are surrounded by scintillator muon 
counters \Ref\cusbnim{Schamberger, R. D. {\it et al.}, NIM
{\bf A309} 450 (1991).}.
In this period we did a whole series of exploratory experiments,
and one of us (J. L-F) also had a wonderful time re-learning quantum
mechanics from ``guru" K. Gottfried.
 \REF\meter{Yoh, J. K.,
{\it et al.}, High Energy Physics-1980, Durand,  Pondrum Eds., AIP, 
NY 700 (1981).}
We resolved the first three 
$\Upsilon$ resonances to the chime of the CUSB ``R" meter 
(a scalar counting of a restrictive trigger) \refmark{\meter}.
We saw the first $\Upsilon$(4S) signal 
in December 1979 \Ref\fin{Finocchiaro, G. {\it et al.}, 
Phys. Rev. Lett. {\bf 45}, 222 (1980).}.
We observed the $\pi\pi$ transitions between the $\Upsilon$ 
bound states \Ref\pipi{Mageras, G. {\it et al.}, 
Phys. Rev. Lett. {\bf 46}, 1115 
(1981); Phys. Lett. {\bf 118B} 453 (1982).}.
We discovered the $\chi _b$ states from inclusive and exclusive
E1 transitions \Ref\han{Han, K. {\it et al.}, Phys. Rev. 
Lett. {\bf 49}, 453 (1982);
Klopfenstein, C. \etal, Phys. Rev. Lett. {\bf 51}, 160 (1983);
Eigen, G. {\it et al.}, Phys. Rev. Lett. {\bf 49}, 1616 (1982);
Pauss, F. {\it et al.}, Phys. Lett. {\bf 130B}, 1439 (1983).}.
\REF\dan{Peterson, D. {\it et al.}, Phys. Lett. {\bf 114B}, 277 (1982).}
\REF\gia{Gianinni, G. {\it et al.} Nuc. Phys. {\bf B206}, 1 (1982).}
We even foretold the existence of $\chi_b$ transitions from studying
event shapes\refmark{\dan},
studied $K^0$ production at all the 
resonances \refmark{\gia},
and made the first determination of $\alpha_s$ from study of $\gamma gg$
on the $\Upsilon$ resonances\Ref\gamgg{Schamberger, R. D., Phys. Lett.
{\bf 138B}, 225 (1984).}.
\REF\semi{Spencer, L. \etal, Phys. Rev. Lett. {\bf 47}, 771 (1981); 
Klopfenstein, C. \etal, Phys. Lett. {\bf 130B}, 444 (1983); 
Levman, G. \etal, Phys. Lett. {\bf 141B}, 271 (1984).}
From seeing high energy electrons emerging from 4S
events we inferred the B-mass, plotted out the $B$ meson semileptonic 
decay spectra \refmark{\semi},
and had long discussions with phenomenologists \Ref\accmm{Altarelli, G.
\etal, Nuc. Phys. B {\bf 208} 365 (1982).}.
\eject

\vbox{\centerline{\epsfig{file=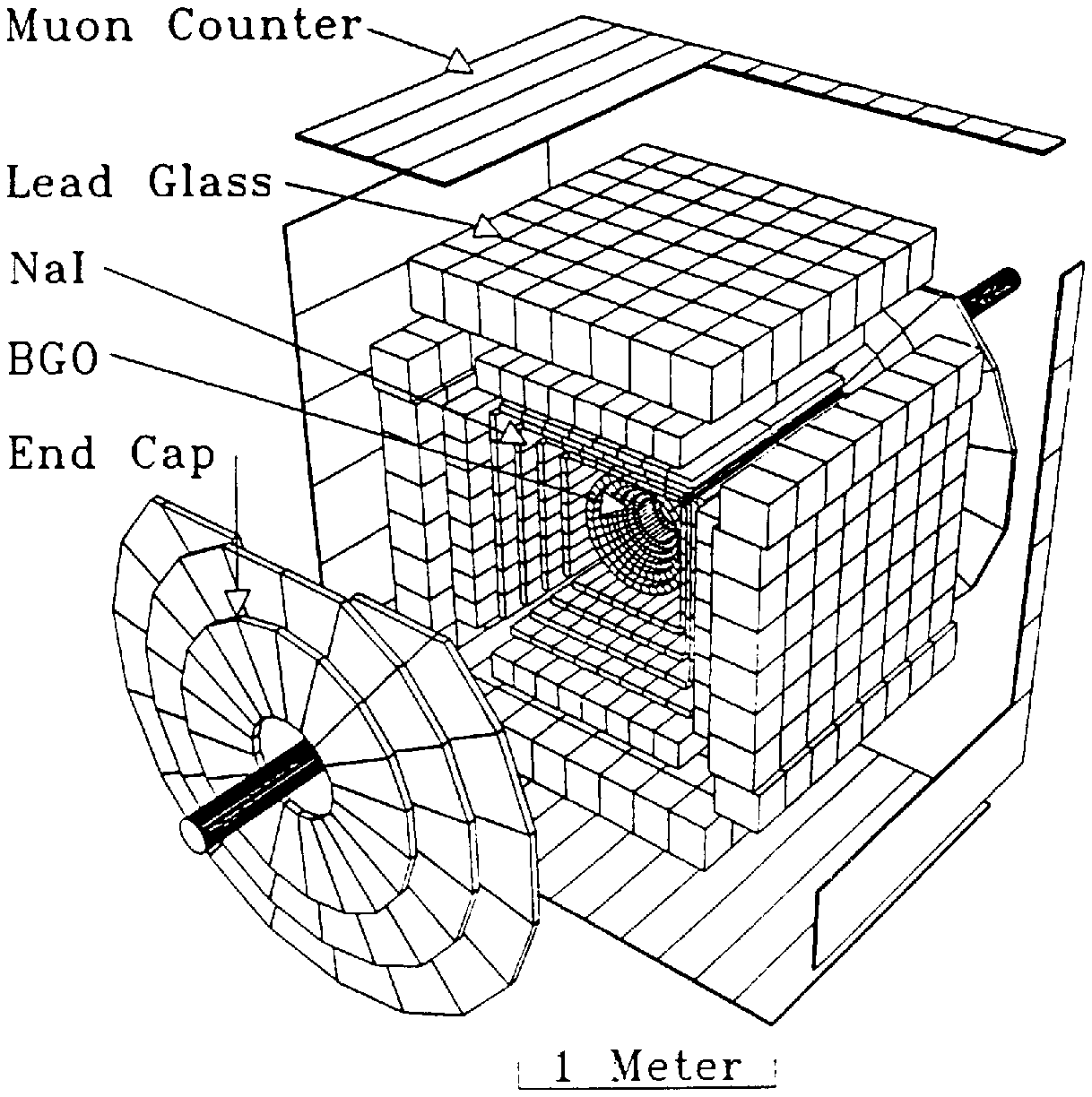,width=10.4truecm}}}  
\vglue2mm
\vbox{\parshape=1 6mm 136.5mm 
\tenpoint\baselineskip10pt
\noindent{\bf Figure \cusb.} The complete CUSB detector. This 
configuration was only reached in 1989. The BGO calorimeter and the 
endcap trigger counters were not 
present at the beginning. Because of the high cost of NaI, lead glass 
was used as a catcher for the electromagnetic showers. The lead glass 
was used later for muon identification.}

\vglue2mm
\vbox{\centerline{\epsfig{file=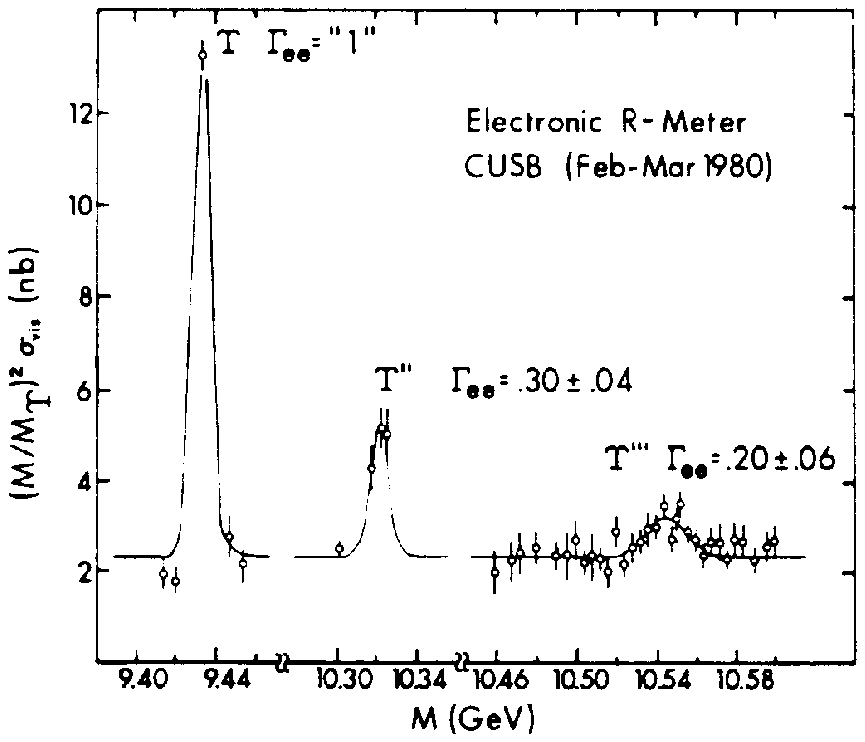,width=9.4truecm}}}  
\vbox{\parshape=1 6mm 136.5mm 
\tenpoint\baselineskip10pt
\noindent{\bf Figure \rmeter.} The \Up4\ was observed by CUSB by counting 
on scalers small angle Bhabha and hadronic \Y+continuum events. Thus the 
fourth upsilon was discovered without analysis of the collected events. 
The latter confirmed the former and gave information on event shape and 
on the production of $B$ mesons.}
\endpage

We searched for the $B^*$ repeatedly on
the $\Upsilon$(4S), in vain, finally observing it as a broadish bump in the 
continuum above (due to the Doppler shift of the 
parent $B$'s) \Ref\bstar{Schamberger, R. D. \etal, 
Phys. Rev. D {\bf 26}, 720 (1982); Phys. Rev. D{\bf 30}, 1985 (1984).}.
This last signal inspired us to improve our photon resolution by
using BGO crystals in addition to NaI ones.
We saw a rise in $\Delta R$ of $\simeq$1/3 due to the production
of \bbar-quarks \Ref\rice{Rice, E. \etal, 
Phys. Rev. Lett. {\bf 48}, 906 (1982).},
and did a complete coupled channel analysis to obtain the parameters
of the higher $\Upsilon$ resonances \Ref\lovelock{Lovelock D. M. J.
\etal, Phys. Rev. Lett. {\bf 54}, 377 (1985).}.
In July 20, 1984 we inserted a BGO quadrant inside the CUSB-I NaI array.
With this set up, together with the Cornell Polarization group,
we made a precision $\Upsilon$ mass measurement \Ref\Mackay{Mackay, W. W.
\etal\ and the CUSB Collaboration, Phys. Rev. D {\bf 29}, 2483 (1984).}.
We also continued to establish limits for non existent particles 
such as axions, short lived particles, light gluinos, $\zeta$ (8.3),
light Higgs, and light squarks \Ref\limits{Sivertz, M. \etal, 
Phys. Rev. D {\bf 26}, 717 (1982); Mageras, G. \etal, Phys. Rev. 
Lett. {\bf 56}, 2672 (1986); Tuts, P. M. \etal, Proc. of the DPF-APS 
1984 Meeting, M. Nieto and T. Goldman Eds., AIP, New York (1985);
Phys. Lett. {\bf B186}, 233 (1987);
Franzini, P. \etal, Phys. Rev. D {\bf 35}, 2883 (1987).}.

In Dec. 1985 we installed the complete BGO array inside the CUSB-I 
detector which is the heart of CUSB-II. The array is a cylinder composed 
of 360 trapezoid cross sectioned BGO crystals, the
5 layers's thickness was twelve radiation length.

\vglue12mm
\vbox{\centerline{\epsfig{file=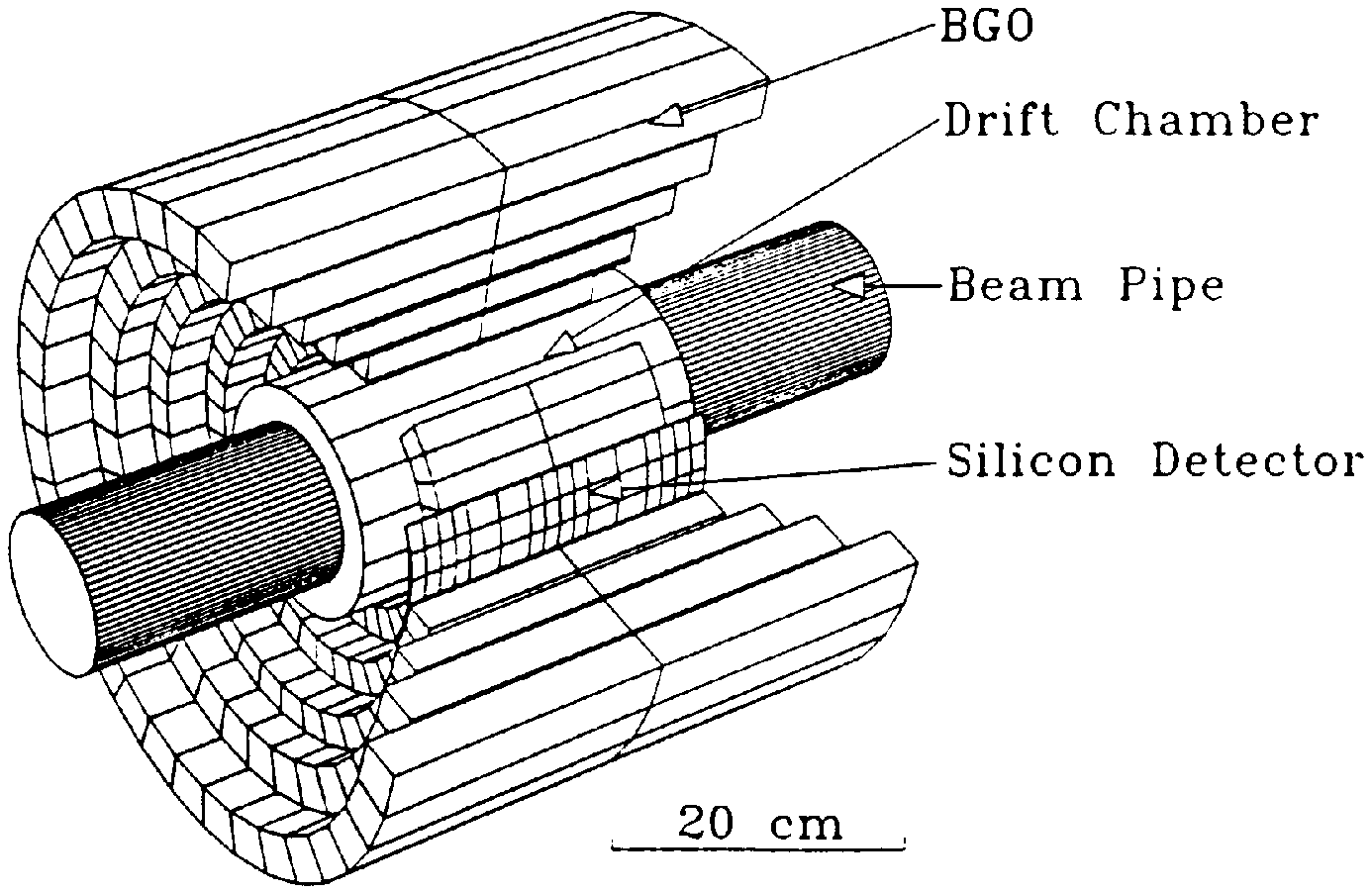,height=8.9truecm}}}  
\vglue3mm
\vbox{\parshape=1 6mm 136.5mm 
\tenpoint\baselineskip10pt
\noindent{\bf Figure \bgo.} The ``BGO cylinder'' improved the CUSB 
energy resolution for low energy photons by a factor of two. In its final 
configuration, silicon pads were between BGO crystals to obtain better angular 
resolution for photons. A small drift chamber inside the BGO array 
detected charged particles.}
\vglue2mm

\noindent
We proved by study of $\approx$5 GeV electrons from Bhabha scatterings 
that the new CUSB-II detector had achieved an improvement of two in 
resolution \Ref\jlfnim{Lee-Franzini, J., NIM {\bf A263} 35 (1988).}.
The CUSB-II physics results obtained in this configuration 
include accurate measurements of the branching ratios 
$B_{\mu\mu}$(3S) and $B_{\mu\mu}$(1S)\Ref\bmumu{Kaarsberg, T. \etal, 
Phys. Rev. D {\bf 35}, 2265 (1987); Phys. Rev. Lett. {\bf 62} 2077 (1989).},
semileptonic branching ratios from $B_u$-mesons and $B_s$-mesons 
semileptonic decays at the $\Upsilon$(4S) and 
$\Upsilon$(5S)\Ref\chiaki{Yanagisawa, C. \etal, Phys. 
Rev. Lett. {\bf 66} 2463 (1991).},
stringent limits on direct photon production from the 
$\Upsilon$(4S)\Ref\direct{Narain, M. \etal, Phys. 
Rev. Lett. {\bf 65} 2749 (1990).},
precise determinations of hyperfine splittings in the $B$-meson system, 
and evidence for $B_s$-meson production on the $\Upsilon$(5S) 
resonance\Ref\jhfs{Lee-Franzini, J. \etal, Phys. 
Rev. Lett. {\bf 65} 2947 (1990).}.
From April 23, 1988 to October, 1989, during CESR's shutdown for 
completing CLEO-II, we constructed and installed the remaining portions 
of CUSB-II, including new BGO read-out electronics, a shower centroid detector,
and a forward-backward lepton end-cap trigger. 
Data obtained from the last CUSB run, Oct. 20, 1989 to Oct. 8, 1990, 
resulted in precision measurements of the relative contributions 
of the spin-orbit and tensor interactions to the fine structure of 
the 2P state and we determined that the long-range
confining potential transforms as a Lorentz 
scalar\Ref\finest{Narain, M. \etal, Phys. 
Rev. Lett. {\bf 66} 3113 (1991).}.
We made precise studies of the sequential decays of the $\Upsilon$(3S),
measured the hadronic widths of the $\chi_b'$ states and
observed the rare decay transition from 
the $\Upsilon$(3S) to the $\chi _b$\Ref\upthrees{Heintz, U. \etal, 
Phys. Rev. Lett. {\bf 66} 1563 (1991);
Phys. Rev. D {\bf 46} 1928 (1992).}.
We measured the $\pi\pi$ mass spectra of the dipions from
$\Upsilon$(3S) hadronic transitions to (2S) and (1S) and
determined the beam energy window where single
$B^*B$ production is dominant\Ref\wu{Wu, Q. W. \etal, Phys. Lett. 
{\bf B301} 307 (1993); Phys. Lett. {\bf 273} 177 (1991).}.

\noindent
The total running period for CUSB-I, CUSB I.5, and CUSB-II
span from October 1979 to October 8, 1990.
The CUSB collaboration had always been a small one, with at most two dozen
members at any one time during the CUSB-I stage\Ref\cusbone{
CUSB-I began with members from Columbia University which in total included: 
T. B\"oh\-ring\-er, F. Costantini, J. Dobbins, \undert{P. Franzini}, 
K. Han, S. W. Herb, D. M. Kaplan, L. M. Lederman, G. Mageras, 
D. Peterson, E. Rice, D. Son, J. K. Yoh, S. Youssef, and T. Zhao,
and similarly, from SUNY at Stony Brook: G. Finocchiaro, G. Gianinni, 
J. E. Horstkotte, D. M. J. Lovelock, C. Klopfenstein, J. Lee-Franzini, 
R. D. Schamberger,  M. Sivertz, L. J. Spencer, and P. M. Tuts. 
We were joined in 1981 by R. Imlay, G. Levman, W. Metcalf and V.
Sreedhar of Louisiana State Univesity and G. Blanar, H. Dietl, G. Eigen,
V. Fonseca, E. Lorenz, F. Pauss, L. Romero and H. Vogel of MPI Munich. 
Louisiana State and MPI Munich left in 1984.}\ 
and no more than one dozen at any one time, including students,
technicians and senior physicists for CUSB-II\Ref 
\cusbtwo{CUSB-I.5, CUSB-II: \undert{P. Franzini}, U.
Heintz, J. Horskotte, T. Kaarsberg, S. Kanekal, \undert{J. Lee-}
\undert{Franzini}, 
D. M. J. Lovelock, M. Narain, R. D. Schamberger, S. Sontz,  P. M. Tuts, 
J. Willins, Q.W. Wu, C. Yanagisawa, S. Youssef, and T. Zhao.}. 
The following table gives a summary of the run history and physics
highlights of each run period.

\vglue 5mm
\vbox to 13cm{
\vbox{\parshape=1 13.25mm 122mm 
\tenpoint\baselineskip11pt
\noindent{\bf Table I.} The data taking history of CUSB is presented in 
terms of the actual runs. For each the integrated luminosity and the 
number of collected events is listed. Some major physics topic studied 
are also listed.}
\vglue-5mm
$$\vbox{\tenpoint\baselineskip10pt
\halign{\hfil#\ &\hfil#\ &\hfil#&\quad#\hfil\cr
\noalign{\vskip1pt\hrule\vskip1pt}
\noalign{\vskip1pt\hrule\vskip1mm}
\multispan4 Run 394-5513\quad Oct. 18, 1979 -- Jan 25, 1984
\quad CUSB\hfill \cr
&${\cal L}$ (pb$^{-1}$)&Events&Physics\cr
\noalign{\vskip1mm\hrule\vskip1mm}
1S & 7.1 & 111,405&Resonances first seen in ``R-meter" 1979.\cr
2S & 29.6 & 224,163&($\pi\pi$, E1) transitions to 1S, discovery of $\chi _b$.\cr
3S & 17.7 & 89,096&3S '79, ($\pi\pi$, E1) to (2S)1S, discovery of $\chi _b'$.\cr
4S & 63.1 & 190,280& 1st seen in Dec. 1979, B semileptonic decay, no $B^*$.\cr
cont & 32.7 & 72,865&$\Delta R\simeq1/3$.\cr
$>$4S & 113.8 & 260,443&Coupled channel $\Upsilon(5S,6S,7S)$, observe
$B^*$'s.\cr
\noalign{\vskip1mm\hrule\vskip1mm}
     & 263.7 & 948,252&$K^0$ production measured, no axion seen.\cr
\noalign{\vglue 4mm}
\multispan4 Run 5515-7284\quad Jul. 20, 1984 -- Aug. 2, 1985\quad CUSB-1.5
\hfill\cr
&${\cal L}$ (pb$^{-1}$)&Events&Physics\cr
\noalign{\vskip1mm\hrule\vskip1mm}
1S & 28.9 & 529,537&Precision (1S) mass (10ppm), $\gamma gg$\cr
4S & 60.3 & 243,916&Semileptonic $B\to\mu ,e, \nu$ spectra, no $b \to u$. \cr
cont. & 23.9 & 70,283&No $B^*$'s here (below 4S), nor on 4S.\cr
\noalign{\vskip1mm\hrule\vskip1mm}
         & 113.3 & 844,599&No $\zeta(8.3)$.\cr
\noalign{\vglue 4mm}
\multispan4 Run 10391-12210\quad Dec. 16, 1985 -- Apr. 23, 1988
\quad CUSB-II\hfill\cr
&${\cal L}$ (pb$^{-1}$)&Events&Physics\cr
\noalign{\vskip1mm\hrule\vskip1mm}
1S & 24.4 & 458,492&$B_{\mu\mu}(1S)$.\cr
3S & 144.3 & 1,001,329&$B_{\mu\mu}(3S)$, no $\eta_b$ seen, no $h_b$ seen.\cr
4S & 273.7 & 1,163,213&Precise (4S) semileptonic decay studies.\cr
cont & 122.3 & 395,013&No $B^*$'s below 4S.\cr
$>4S$ & 140.2 & 467,729&HFS, $B^*, B_s, B^*_s$ masses, $B\to e, \nu$ at 5S. \cr
\noalign{\vskip1mm\hrule\vskip1mm}
    & 704.9 & 3,090,763& Limits: Higgs, squark, gluino, short $\tau$
particles.\cr  } }$$
\vfill}
\endpage

\vbox to 5.5cm{
\vbox{\parshape=1 13.25mm 122mm 
\tenpoint\baselineskip11pt
\noindent{\bf Table I.} Continued.}
\vglue-5mm
$$\vbox{\tenpoint\baselineskip11pt
\halign{\hfil#\ &\hfil#\ &\hfil#&\quad#\hfil\cr
\noalign{\vskip1pt\hrule\vskip1pt}
\noalign{\vskip1pt\hrule\vskip1mm}
\multispan4 Run 20001-22542\quad Oct. 20 1989 -- Oct. 8, 1990
\quad CUSB-II+\hfill\cr
&${\cal L}$ (pb$^{-1}$)&Events&Physics\cr
\noalign{\vskip1mm\hrule\vskip1mm}
1S & 18.5 & 332,413&$\alpha_s(1S)$, hadronic width.\cr
3S & 143.6 & 969,8903&$\alpha_s(3S)$, hadronic width, precision FS, 
$\pi\pi$ spectra.\hglue1mm\cr
4S & 36.9 & 150,750&No direct photons from 4S, no $\pi\pi$ to (2S), (1S).\cr
cont & 32.4 & 99,608&   \cr
\hglue1mm$BB^*$ & 63.1 & 204,356&Precision $B^*$ mass.\cr
\noalign{\vskip1mm\hrule\vskip1mm}
              & 294.5 & 1,757,617&$\Lambda_{\overline{MS}}$.\cr
\noalign{\vglue4mm}
\multispan4 Run 394-22542\quad Oct. 18 1979 -- Oct. 8, 1990
\quad CUSB to CUSB-II+\hfill\cr
&${\cal L}$ (pb$^{-1}$)&Events&Physics\cr
\noalign{\vskip1mm\hrule\vskip1mm}
 All   & 1376.4 & 6,641,231&Scalar nature of confining potential.\cr} }$$
\vfill}

\twelvepoint
\vglue10mm
\cl{\fourteenpoint\bf SELECTED PHYSICS RESULTS}
\vglue3mm

\vglue2mm
\cl{\bf $\Lambda_{\overline{MS}}$ from $\Upsilon$'s $\to\mu\mu$}
\vglue2mm

\def\bmm{\ifm{B_{\mu\mu}}} \def\Gam{\ifm{\Gamma}}
\def\als{\ifm{\alpha _s(m_b)}} \def\lams{\ifm{\Lambda_{\overline{MS}}}}
From our measurements of the branching ratio for \Y\to\muu, \bmm, and 
the other measured parameters of the \Y's we obtain the branching 
ratio for \Y\to$gg$ and thus we determine \als\ and \lams.
\vglue4mm
\vbox to3cm{
\vbox{\parshape=1 24.25mm 100mm 
\tenpoint\baselineskip10pt
\noindent{\bf Table II.} Measurements of \bmm\ and the derived \als\ 
values.}
\vglue-5mm
$$\vbox{
\tenpoint\baselineskip10pt
\tabskip=5mm
\halign{\hfil#\hfil&\hfil#\hfil&\hfil#\hfil&\hfil#\hfil&\hfil#\hfil\cr
\noalign{\vskip1pt\hrule\vskip1pt}
\noalign{\vskip1pt\hrule\vskip1mm}
Resonance&\bmm\ (\%)& \Gam\ (keV) & \als\          & \lams \cr
\Up1 & 2.61\PM 0.09 & 51.1\PM 3.2 & 0.174\PM 0.004 & 150\PM 13 \cr
\Up2 & 1.38\PM 0.25 & 42.3\PM 9.2 & 0.176\PM 0.016 & 167\PM 58 \cr
\Up3 & 1.73\PM 0.15 & 27.7\PM 3.7 & 0.173\PM 0.008 & 154\PM 29 \cr
\noalign{\vskip1pt\hrule\vskip1mm}
Average & --- & --- & 0.1736\PM 0.0033\PM 0.017 & 157\PM 12\PM 60 \cr
}}$$
\vfill}
\noindent
For the average values of $\alpha _s(m_b)$ and $\Lambda_{\overline{MS}}$
we have included a reasonable guess of the theoretical uncertainty in fixing
the energy scale in the systematical error.
The $\alpha _s$ and $\Lambda_{\overline{MS}}$
obtained by us are in excellent agreement with those obtained using a 
number of other processes, proving that the $\Up$ system 
provides an independent probe of QCD.

\vglue2mm
\cl{\bf Hyperfine Splitting of \B\ and \BS\ Mesons} 
\vglue2mm

\noindent
We have studied the inclusive photon
spectrum from $2.9\times 10^4$  $\Upsilon$(5S) decays.
We observe a strong signal due to \Bs\to\B\gam\ decays,
both from inclusive hadronic events a) and from electron
tagged events b). From a detailed analysis we obtain: i) the
average \Bs--\B\ mass difference, ($46.7\pm0.4$) MeV, ii) the photon 
yield per  $\Upsilon$(5S) decay, $\langle \gamma/\Uvs\rangle=1.09\pm 0.06$ 
and iii) the average velocity of the \Bs's, $\langle \beta \rangle=0.156 \pm
0.010$, for a mix of non strange (\B) and strange (\BS) \Bs--mesons 
from  $\Upsilon$(5S) decays. From the shape of the photon line, we find that 
significant production of \BS\ is required implying nearly equal 
values for the hyperfine splitting of the \B\ and \BS\ meson systems.

\vbox{\centerline{\epsfig{file=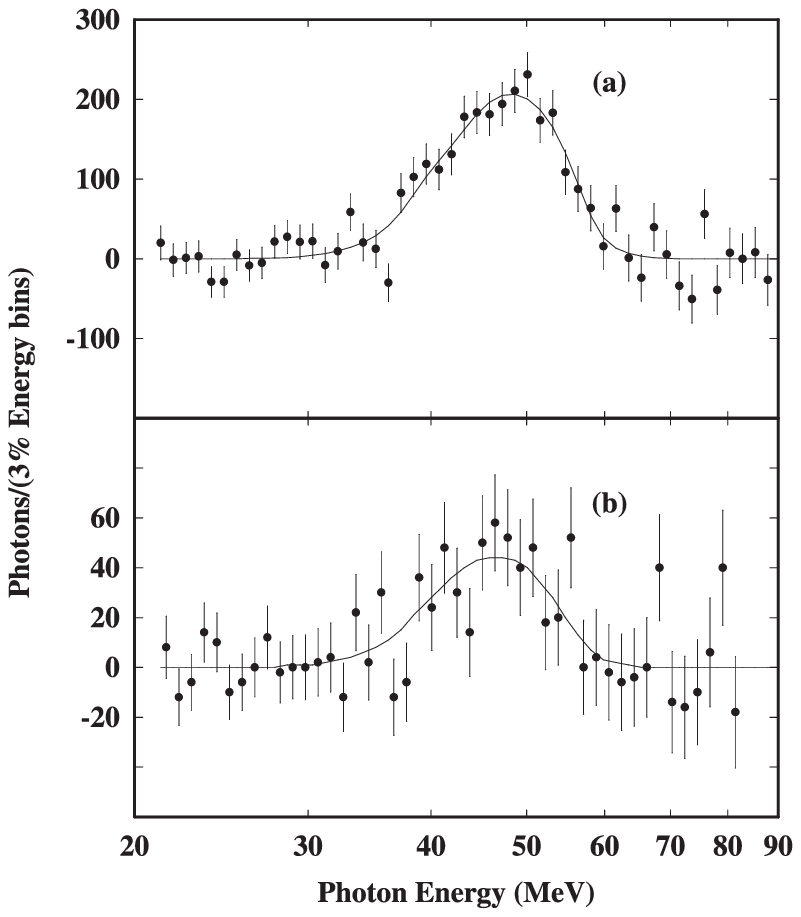,height=10truecm}}}  
\vbox{\parshape=1 6mm 136.5mm 
\tenpoint\baselineskip10pt
\noindent{\bf Figure \bstarb.} 
Photon spectrum in inclusive hadronic events above the flavor threshold.
(a) The line at  \ab48 MeV is due to the decay $B^*$\to$B$+\gam, proving
the existence of the $B^*$ meson and measuring its mass. (b) The line is
present also selecting events with a high energy electron, confirming 
the presence of a  $B$ meson decaying semileptonically.} 

\vglue8mm
\cl{\bf $B$ Semileptonic Decays at the $\Upsilon$(4S) and 
the $\Upsilon$(5S)}
\vglue2mm

\noindent
$B$ meson semileptonic decay spectra have been obtained 
at the $\Upsilon$(4S) and at the $\Upsilon$(5S).
The branching ratio for $B\to e\nu X$ at the
$\Upsilon$(4S) is found to be $(10.0\pm0.5)\%$. 
The electron spectrum of $B\to e\nu X$ at the $\Upsilon$(5S) is 
observed for the first time and the average branching ratio for 
$B,B_s\to e\nu X$ is consistent with that for $B$'s from $\Upsilon$(4S) 
decays. The shape of the electron spectrum at the $\Upsilon$(5S) 
indicates production of $B$ mesons which are
heavier than non-strange $B$'s, presumably $B_s$'s.

\vglue6mm
\cl{\bf \bbar\ Spectroscopy from the $\Upsilon$(3S) State}
\vglue2mm

\vbox{\centerline{\epsfig{file=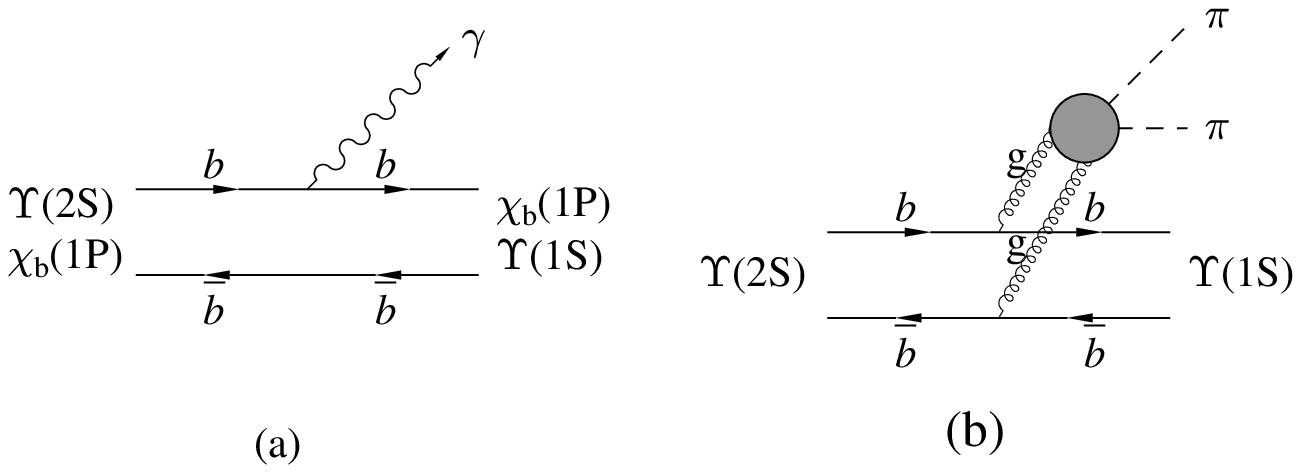,width=9truecm}}} 
\vbox{\parshape=1 6mm 136.5mm 
\tenpoint\baselineskip10pt
\noindent{\bf Figure \eone.} Amplitudes for: (a) electric dipole 
transitions (E1) \up3$S$(\up3$P$)\to\up3$P$(\up3$S$)+\gam\
and (b) double {\it color-electric} dipole transitions
$n\,$\up3$S$\to$(n-1)\,$\up3$S$+$gg (2\pi)$.}

\vbox{
\vbox{\centerline{\epsfig{file=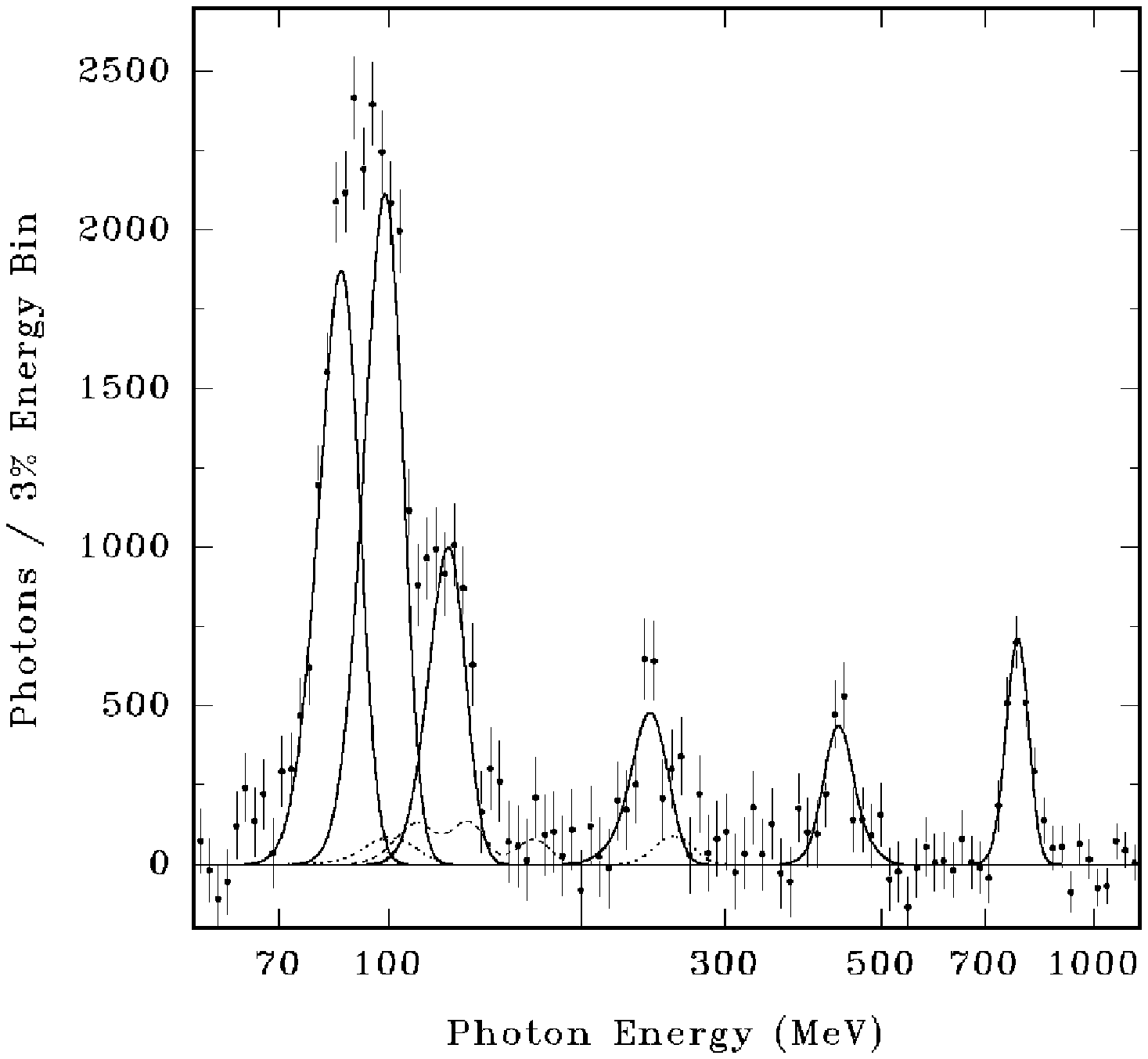,width=105mm}}}  
\vglue3mm
\vbox{\parshape=1 6mm 136.5mm 
\tenpoint\baselineskip10pt
\noindent{\bf Figure \gamspec.} The inclusive photon spectrum at the 
\Up3, after background subtraction. Both lines from direct \Up3 decays as 
well as from decays of daughter \bbar\ states, see figure 1, are 
observed.}
}

\noindent
We have made a detailed study of hadronic and electric dipole (E1) transitions
between the \Ups\ and \chib\ states, both in exclusive and inclusive channels.
We have determined their branching ratios: $\BR(\Ypp \to\xb(2P_{2,1,0})
\gam)$ = (11.1\PM0.5\PM0.4)\%, (11.5\PM0.5\PM0.5)\%, (6.0\PM0.4\PM0.6)\%. We 
have measured the center of gravity of the \xb(2P) states to be 
(10259.5\PM0.4\PM1.0) MeV. We have made precision measurements of the 
electric dipole transition rates from $\Ypp$ to $\Xp$, they are in 
excellent agreement with theory.
\vglue1mm
\vbox{\centerline{\epsfig{file=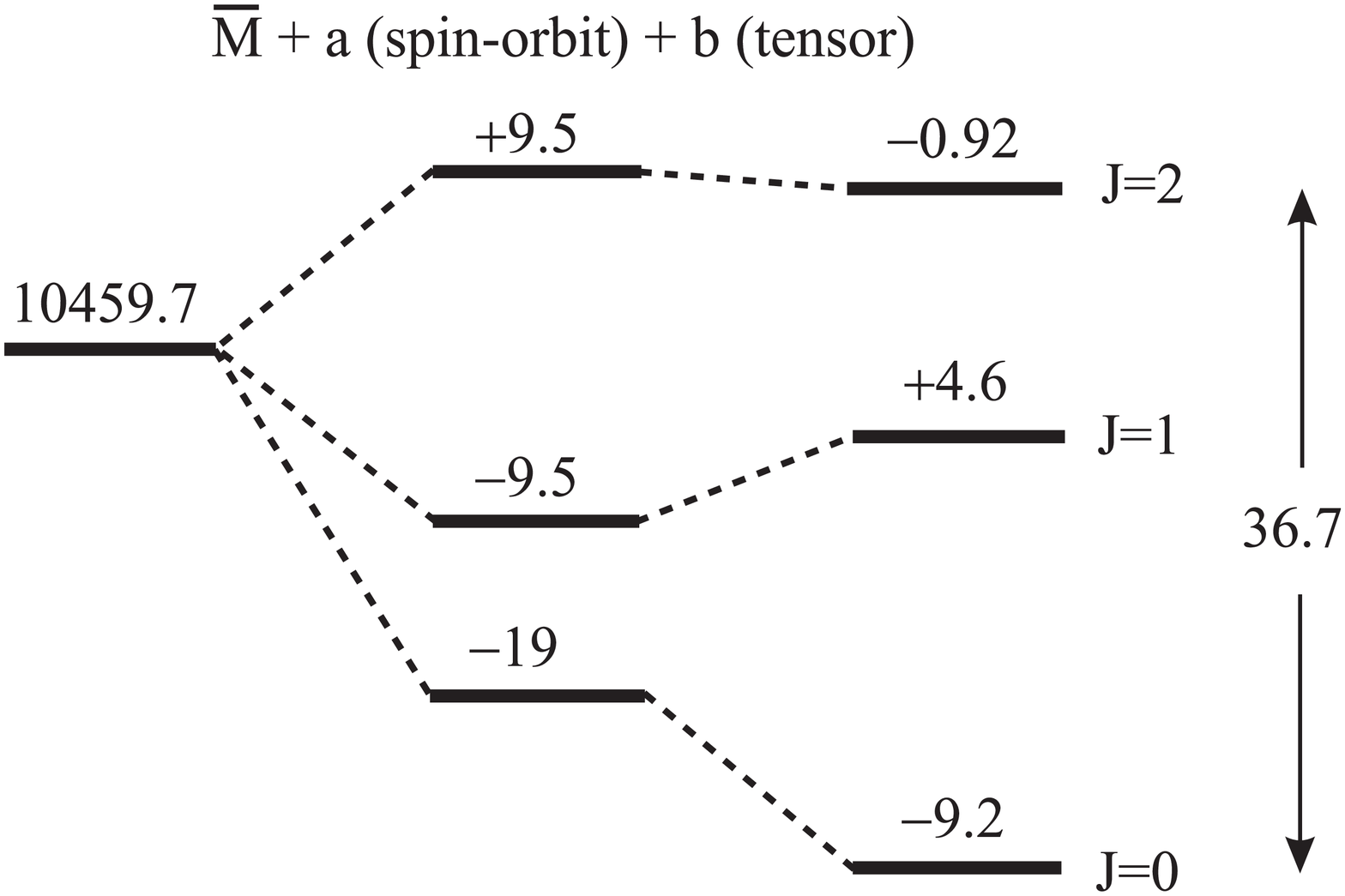,width=7.5truecm}}}  
\vglue 3mm
\vbox{\parshape=1 6mm 136.5mm 
\tenpoint\baselineskip10pt
\noindent{\bf Figure \FineS.} Fine structure of the $\chi_b''$ state, 
resolved in their spin-orbit and tensor contributions as determined by 
CUSB. Energies are in MeV.}

{\def\Xp{\chi}
\noindent
The fine structure splittings obtained using all data is 
$M(\Xp_2)-M(\Xp_1)=(13.5\pm 0.4\pm 0.5)$ MeV and
$M(\Xp_1)-M(\Xp_0)= (23.2\pm 0.7\pm 0.7)$ MeV, leading to a ratio}
\endpage

\vglue42mm
\vglue2mm
\vbox{
\vbox{\centerline{\epsfig{file=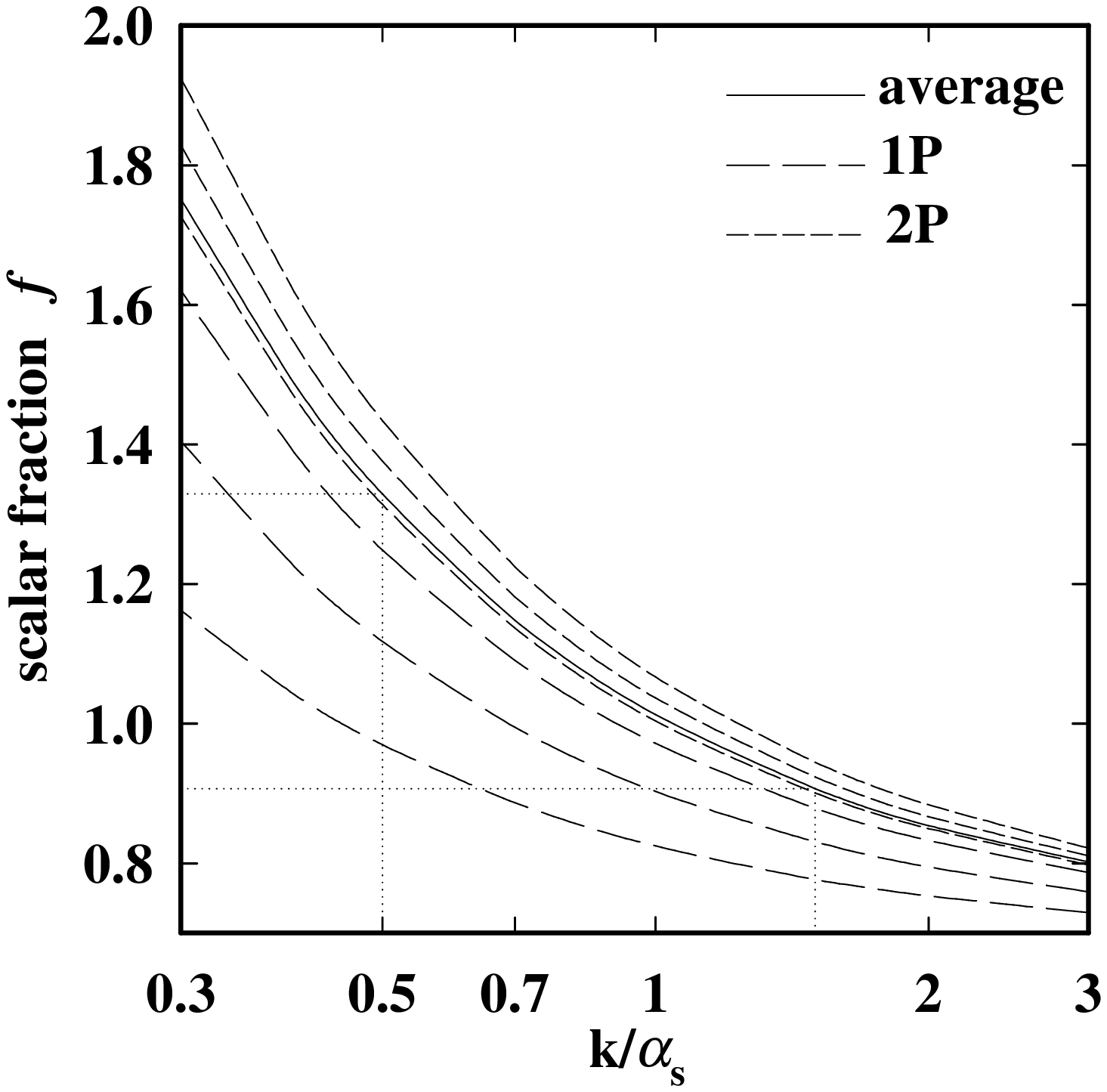,height=105mm}}}  
\vbox{\parshape=1 6mm 136.5mm 
\tenpoint\baselineskip10pt
\noindent{\bf Figure \Fscal.} Scalar contribution to the longe range 
confining potential for \X, \Xp\ and their average as function of 
$\alpha_s$ obtained using the fine structure measurement from CUSB.}
\vglue2mm}
\vglue-165mm\baselineskip13pt\noindent
$R=0.584\pm0.024\pm0.02$. The fine structure measures the relative 
contributions of the spin-orbit interaction $a=(9.5\pm 0.2\pm 0.1)$ MeV 
and tensor interaction $b=(2.3\pm 0.1\pm 0.1)$ MeV,
that is, the spin-orbit interaction dominates over the 
tensor interaction.
We also find that the long-range
confining potential is due to the exchange
of an effective Lorentz scalar. 
From the measured branching ratios we infer the hadronic widths of the
$\xb(2P)$ states and find them to be consistent with QCD predictions. We
use them to derive values of $\alpha_s$. We have also observed the
suppressed transition \Up3\to\xb$(1P)$\gam. The measured
branching ratio suggests that relativistic effects are important. 
\vglue125mm

\vglue2mm
\cl{\bf Study of \pic\ Transitions from the \Up3\ State}
\vglue2mm

\noindent
We have investigated the decay $\Upsilon(3S)$\to$\Upsilon(1S$, $2S)\pic$,
where the final state $\Upsilon(1S, 2S)$ decays to a pair of leptons. We found 
$\sim$ 390 events of the type $\Upsilon$(3S)\to\break $\Upsilon$(1S)$\pic$ 
and $\sim$ 140 events of the type $\Upsilon$(3S)\to$\Upsilon$(2S)$\pic$.  
The corresponding branching ratios are
$(3.27 \pm 0.30) \%$ and $(3.59 \pm 0.49) \%$ respectively.
We have also studyed the $\pi\pi$ invariant mass spectrum. There have been 
contradictions on the mass spectrum between previous experimental data and
theoretical predictions. We have verified the unusual double-peak behavior on
the dipion mass spectrum from $\Upsilon$(3S)\to$\Upsilon$(2S)$\pic$,
it is quite different from the
spectra for $\pi\pi$ decays from other $\Upsilon$ and $\psi$ 
states as well as theories which predict a single peak in the high 
mass region of the
distribution. We have compared our spectrum with several current theoretical 
modifications to various models and found that none of them could successfully 
explain the observed shape of the double-hump spectrum\Ref\bdm{Belanger, P.,
DeGrand, T. and Moxhay, P., Phys. Rev. D {\bf 39}, 257 (1988).}.

\vskip5mm
\vglue5mm
\vbox{\centerline{\epsfig{file=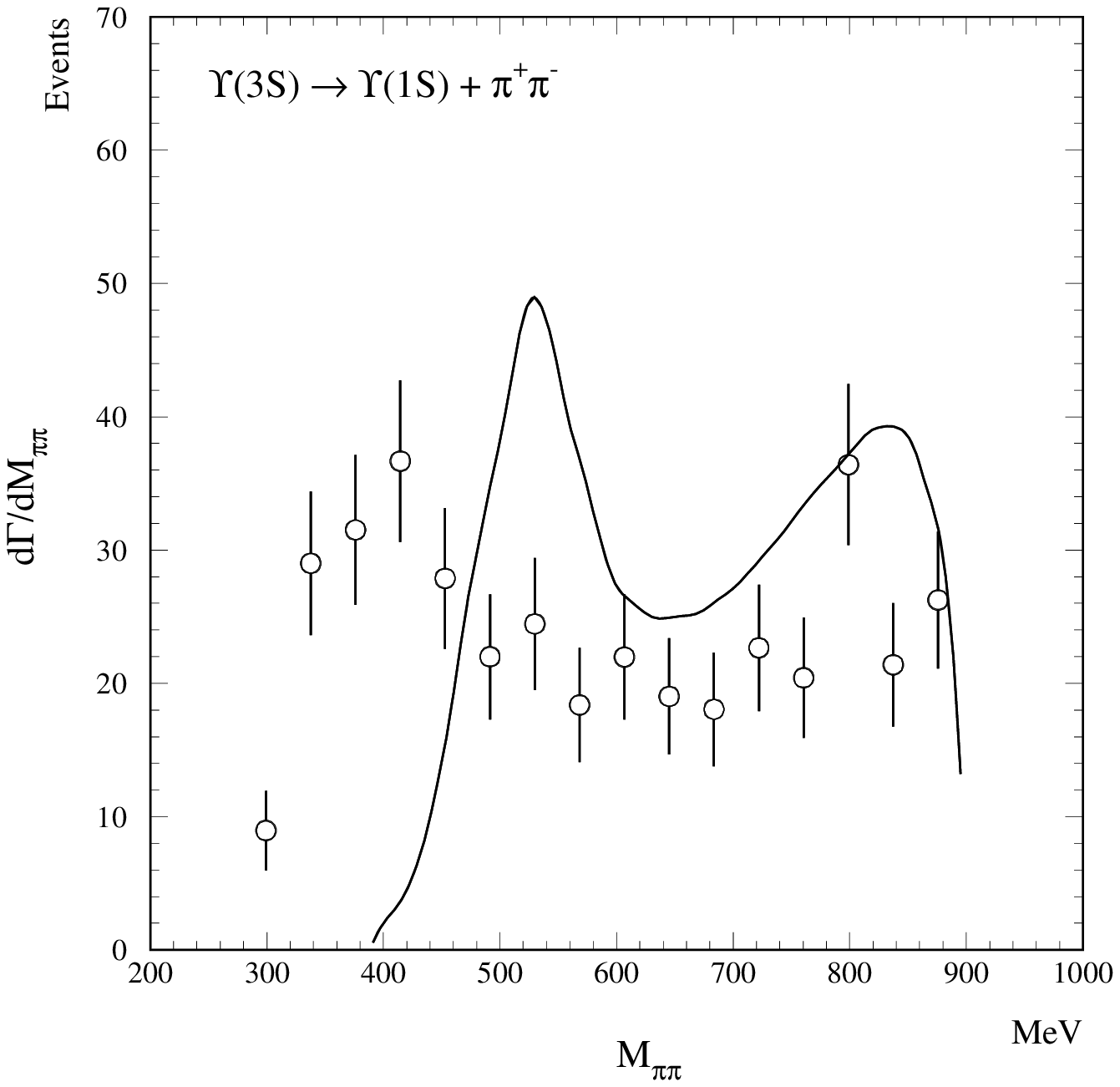,width=12truecm}}} 
\vbox{\parshape=1 6mm 136.5mm 
\tenpoint\baselineskip10pt
\noindent{\bf Figure \dipion.} Dipion mass spectrum observed in 
\Up3\to\Up2+\pic(\pio). The curve is from the model ref. \bdm.}

\vglue6mm
\cl{\fourteenpoint\bf REGRETS}
\vglue3mm
\noindent
While we recall our venture of CUSB at CESR with great affection
and sense of accomplishments, there are regrets that we could not have
stayed, ran on the $\Upsilon$(3S) longer, on the $\Upsilon$(2S) especially 
and found a few more known-to-exist states, such as the $\eta _b$ and 
$h_b$ \refmark{\pafs, \pasv}. We could have
solved a few more puzzles such as the hadronic widths of the 
$\chi _b$ states \refmark{\pajhad} and whether there is a pseudoscalar 
field in the $Q\bar Q$ potential \refmark{\pasv}. There are,
however, other problems which no amount of additional CUSB running
would have helped. For example, there are no light Higgs, Sparticles,
and after we ruled out that there are neither direct photons nor 
excessive direct pions from the $\Upsilon$(4S), there are still no 
explanations why the the $B$-semileptonic decay branching ratios do not agree
with theoretical expectations \refmark{\direct, \wuthesis}.

\break

\vglue3mm
\cl{\fourteenpoint\bf ACKNOWLEDGEMENTS}
\vglue3mm

\noindent
The author wishes to thank Dan Kaplan and the other organizers
for both an extremely enjoyable and a very scholarly reunion of the
beauty physics community. She thanks Dean Schamberger and especially
Paolo Franzini for retrieving archaic CUSB figures so they can
have a new life here and (PF) for editing the manuscript.

   \immediate\closeout\referencewrite
   \referenceopenfalse
\vglue4mm
   \cl{\fourteenpoint\bf REFERENCES}
\vglue4mm
\tenpoint\baselineskip12pt
   \input referenc.texauxil 

\bye